\documentclass[noeprint,twocolumn,showpacs,amsmath,amssymb,sort&compress,floatfix,superscriptaddress,pre,aps,longbibliography]{revtex4-2}
\usepackage{bm}
\usepackage{graphicx, xcolor}
\usepackage{dcolumn}
\usepackage{comment}
\usepackage{siunitx}
\usepackage{amsmath,amssymb,amsfonts}
\usepackage{booktabs}
\usepackage{hyperref}

\newcommand{\rv}[1]{{\textcolor{black}{#1}}}
\newcommand{\rvv}[1]{{\textcolor{black}{#1}}}

\begin{document}

\title{A Proliferating Nematic That Collectively Senses an Anisotropic Substrate}

\author{Toshi Parmar}
 \affiliation{Department of Physics, University of California Santa Barbara,  Santa Barbara, CA 93106, USA}
\author{Fridtjof Brauns}
\affiliation{ Kavli Institute for Theoretical Physics, University of California Santa Barbara, Santa Barbara, CA 93106, USA}
\author{Yimin Luo}
\affiliation{ Department of Mechanical Engineering, Yale University, New Haven, CT 06511, USA}
\author{M. Cristina Marchetti}
\affiliation{Department of Physics, University of California Santa Barbara, Santa Barbara, CA 93106, USA}
\affiliation{Interdisciplinary Program in Quantitative Biosciences, University of California Santa Barbara, Santa Barbara, CA 93106, USA}

\date{\today} 

\begin{abstract}
Motivated by recent experiments on growing fibroblasts, we examine the development of nematic order in a colony of elongated cells proliferating on a nematic elastomer substrate.  After sparse seeding, the cells divide and grow into locally ordered, but randomly oriented, domains that then interact with each other and the substrate. Global alignment with the substrate is only achieved above a critical density, suggesting a collective mechanism for the sensing of substrate anisotropy.  The system jams at high density, where both reorientation and proliferation stop. Using a continuum model of a proliferating  nematic liquid crystal, we examine the competition between growth-driven alignment and substrate-driven alignment in controlling the density and structure of the final jammed state. We propose that anisotropic traction forces and the tendency of cells to align perpendicular to the direction of density gradients act in concert to provide a mechanism for collective cell alignment.
\end{abstract}

\maketitle

\section{\label{sec:introduction} Introduction}

\rv{Biological tissues composed of elongated cells often exhibit nematic order where the cells align with each other along their long axis~\cite{Balasubramaniam2022ActiveMorphogenesis}. Such order and its disruption have been shown to control a variety of biological processes, such as directed collective cell migration~\cite{Wu2025CollectivePerspective}, and play a central role in morphogenetic processes in vivo and in vitro~\cite{Heisenberg2013ForcesPatterning, Casale2021GeometricalTissues, Diaz-de-la-Loza2024TheDriver}. An important question is how the interplay between cell-cell interactions (steric repulsion and adhesion) and influences from the environment control nematic order. It is known that alignment of long biopolymers in the extracellular matrix can guide \emph{individual} cell motion through contact guidance of locomotion~\cite{Li2017OnFibroblasts, Duclos2014PerfectCells, Adar2024Environment-StoredRemodeling, Guillamat2025GuidanceSurfaces}, playing an important role in metastatic invasion \cite{Erdogan2017Cancer-associatedFibronectin}. Less explored is the effect of environment anisotropy on \emph{collective} cell patterns and migration, as relevant, for instance, in wound healing~\cite{Talbott2022WoundFibrosis} and morphogenesis~\cite{Balasubramaniam2022ActiveMorphogenesis}. The ability to use a structured environment to direct collective cell organization and migration also has direct relevance to bottom-up tissue engineering and organoid morphogenesis, where substrate patterns could be used to shape tissues by designing nematic texture~\cite{Guillamat2025GuidanceSurfaces, Guillamat2022IntegerMorphogenesis, Endresen2024Actuation3D, Huang2025Cell-SheetForces}. }

\rv{Recent work has demonstrated that microgrooved and patterned substrates can both direct the dynamics of single-cells and organize the collective dynamics of confluent cell monolayers \cite{Lacroix2024EmergenceGuidance}. More surprising is the observation of the behavior of cells cultured on liquid crystal elastomers (LCE) substrates that are topographically flat, but can be prepared in a nematic state, where the substrate fibers are aligned at the molecular scale. Experiments by one of us~\cite{Luo2023Molecular-scaleMonolayers} and others~\cite{Martella2019LiquidAlignment, Skillin2024StiffnessAlignment} have shown that in this case single cells appear insensitive to the molecular-scale orientation of the substrate, but cell collectives are able to align with the direction of broken symmetry of the substrate and to develop nematic order on the scale of the entire tissue. This suggests a new and intrinsically collective mechanism  for sensing molecular-scale substrate anisotropy. In this work we reexamine these experiments and demonstrate a new collective mechanism for nematic cell alignment due to anisotropic friction. In contrast to a previously studied scenario that requires flow alignment \cite{Thijssen2020ActiveParameter}, the mechanism we propose operates via density gradients and growth dynamics and requires neither flow alignment nor active shear stresses.}
 
\rv{We begin with a brief summary of the experiments~\cite{Luo2023Molecular-scaleMonolayers} to highlight the main open questions that our model will address. In these experiments human dermal fibroblasts (hdFs) are seeded at low density on a LCE substrate. Cells then grow and divide along their long axis,  organizing in locally aligned high density domains separated by low density disordered regions. }

\begin{figure*}[t]
    \includegraphics[width=\linewidth]{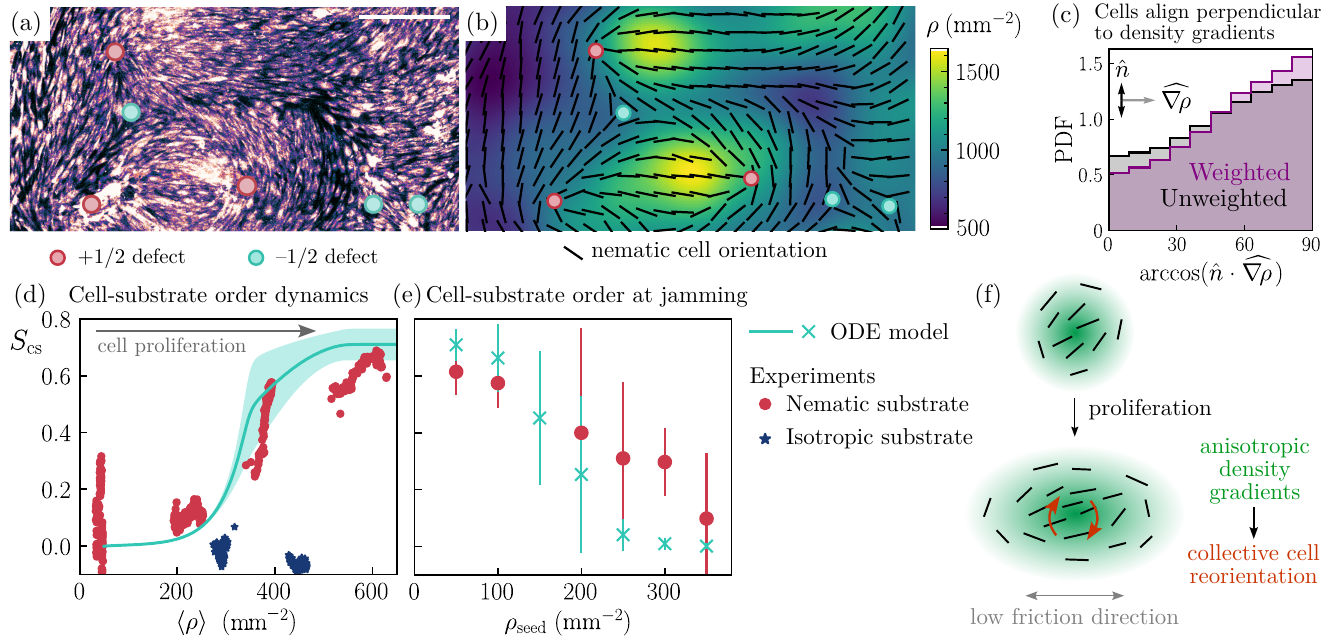}
    \caption{
    \rv{Analysis of experimental data highlights the interplay of order and density. (a--c) Snapshots of (a) jammed hdF cells growing on a LCE substrate, (b) their orientation fields, and the local density (heat-map). Dots in (a) and (b) are topological defects in the orientation field of the cells with charges $+1/2$ (red dots) and $-1/2$ (cyan dots).
    (c) Histogram of the angle between the cell orientation ($\mathbf{\hat{n}}$) and the density gradient ($\nabla\rho$) showing preferential cell orientation perpendicular to density gradients in a jammed monolayer grown on a nematic substrate (data collected from $n=5$ distinct experiments).  The purple histogram is weighted by $S |\nabla\rho|$, while the gray histogram is unweighted.
    (d)~Cell-substrate order parameter in experiments on hdF monolayers growing on a nematic (red dots) and an isotropic (blue stars) substrate. Each group of points corresponds to a separate experiment recorded at different times after seeding with $\rho_\mathrm{seed} = \SI{50}{cells/mm^2}$, with higher densities corresponding to later times as cells proliferate. The cyan solid lines are numerical solutions to Eqs.~(\ref{eq:S}--\ref{eq:rho}) with  parameters: $\gamma_0=0.25$ and $\Pi=0.0015$.
    (e)~Cell-substrate order at jamming as a function of seeding density. Red dots are experimental data from $n=(3, 1, 1, 1, 3, 2)$ experiments respectively with error-bars showing the standard deviation. Cyan crosses represent numerical solutions to Eqs.~(\ref{eq:S}--\ref{eq:rho}) with  parameters: $\gamma_0=0.25$ and $\Pi=0.0015$. The error bars show standard deviation obtained from $50$ simulations with random starting angles.
    (f)~Illustration of the alignment mechanism mediated by the interplay of cell proliferation, anisotropic friction and cell-alignment perpendicular to density gradients.} }
    \label{fig:experiment_fig_1}
\end{figure*}

\rv{While cells are aligned within each domain, the domains themselves are oriented in random directions, } regardless of whether the LCE substrate is in its isotropic or nematic state. Upon a further increase in density, however, the behavior on isotropic and nematic substrates is distinctly different. On isotropic substrates no further alignment across domains is observed as the density continues to increase. On nematic substrates the ordered domains grow in size and align with the direction of order of the substrate, resulting in cell alignment with the substrate at the global scale of the entire monolayer.
At even higher densities cells jam orientationally and eventually growth ceases. \rv{There are two key observations that we aim to elucidate.
First, the degree of nematic order in the jammed state depends on the initial seeding density, with a lower seeding density leading to a more ordered final state. 
Second, individual cells do not align with the direction of liquid crystalline order of the substrate. Moreover, at intermediate density, cells align with each other in large regions via steric effects, but not with the substrate direction. At even higher densities the entire cell layer aligns with the substrate. These observations suggest that substrate alignment occurs via collective sensing.}

\rv{A new analysis of experimental data \cite{Luo2023Molecular-scaleMonolayers} based on a $k$-nearest neighbor (kNN) method has enabled us to tease out the role of density gradients in driving cell alignment. This method allows us to construct continuum fields in systems with strong density fluctuations by averaging over topological instead of metric neighbors~\cite{Biau2015LecturesMethod}. This analysis shows a correlation between cell alignment and the direction of density gradients (see Fig.~\ref{fig:experiment_fig_1}), demonstrating that cells preferentially align tangentially along the contours of high density domains, i.e., perpendicular to the density gradient.  
Guided by this insight, we
formulate a hydrodynamic theory for a proliferating cellular nematic that aligns with external cues and driven towards a jammed state.}

\rv{First we  examine the interplay between alignment due to an external field proportional to cell density and alignment from steric effects in a minimal spatially homogeneous setting. This model provides a simple explanation of the observation that a low initial seeding density results in a more ordered state at long times} as arising from the competition between the rate of cell growth and the rate of cell reorientation that drives nematic order. Essentially, at low seeding density, appreciable order can build up before cells jam due to steric effects.

\rv{We then incorporate spatial inhomogeneities to tease out the mechanisms that drive the collective sensing of substrate order.
Since the LCE substrate is very stiff (${\sim}\, \SI{300}{MPa} $~\cite{Luo2023Molecular-scaleMonolayers}), it is unlikely that mechanical interactions mediated by substrate deformations play a role.
Instead, we propose that nematic order of the substrate enters through anisotropic traction forces, resulting in lower friction along the direction of the substrate's liquid crystalline order. Isotropic pressure due to proliferation of cells then generates anisotropic flows that stretch high density domains along the low friction axis. The resulting anisotropic gradients lead to reorientation of cells, eventually causing their global alignment with the substrate.
We stress that this alignment mechanism is distinct from ones previously considered in the literature, where flows generated by deviatoric active stresses and enhanced by anisotropic friction promote order through flow alignment \cite{Thijssen2020ActiveParameter}
}

\begin{figure}
    \centering
    \includegraphics[width=0.5\linewidth]{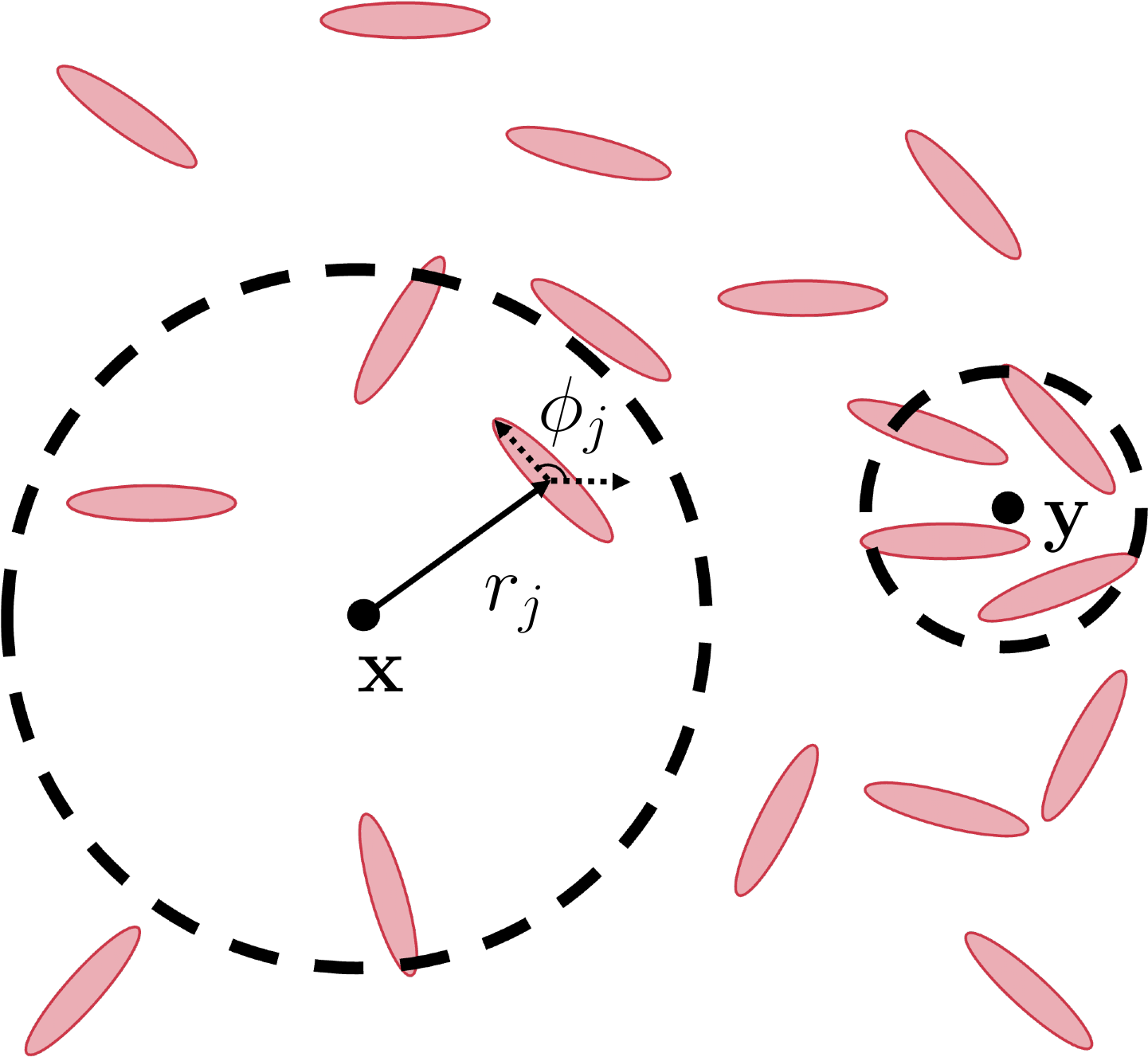}
    \caption{Illustration of kNN coarse-graining method for $k=4$. At point $\mathbf{x}$, where cells are sparse, the coarse-graining radius, chosen to encompass exactly $k$ cells, is larger than at point $\mathbf{y}$ near which cells are dense. Each neighboring cell to point $\mathbf{x}$ is indexed by $j=1,2,...k$ and is at a distance $r_j$ and makes an angle $\phi_j$ with the $x$-axis. For expressions of coarse-grained density and order parameter fields see Sec.~\ref{sec:experiments}.}
    \label{fig:kNN}
\end{figure}
 
The remainder of the article is organized as follows. In Section~\ref{sec:experiments} we present a new analysis of experimental data that demonstrates the correlation between density gradients and cell orientation. In Section~\ref{sec:model} we formulate the continuum model of a proliferating nematic capable of jamming. In Section~\ref{sec:homogeneous} we examine the competition between cell growth and substrate-driven alignment in the case of a spatially homogeneous system. In Section~\ref{sec:inhomogeneous} we consider the spatially extended model and show that anisotropic tractions and density fluctuations together provide a mechanism for collective cell alignment to the substrate. We conclude in Section~\ref{sec:discussion} with a summary of our results and some comments on open questions. Section~\ref{sec:methods} gives details on data analysis and computational methods.

\section{Interplay of Ordering and Density in hdF-LCE System}
\label{sec:experiments}
In the following we describe a new analysis of previously published  experimental data for hdF cells growing on an LCE substrate~\cite{Luo2023Molecular-scaleMonolayers}. To construct continuum fields from discrete experimental data we use a $k$-nearest neighbor (kNN) method, which allows us to describe non-confluent monolayers, with large density fluctuations and empty regions between cells. Let the $k$ cell nuclei nearest to an arbitrary query point $\mathbf{x}$ be labeled by $j=1,2,\cdots, k$. Each of the $k$ elongated nuclei is located at a distance $r_j$ from $\mathbf{x}$ and oriented along the angle $\phi_j$ measured from the positive $x$-axis. For a graphical illustration of the method see Fig.~\ref{fig:kNN}. 
The coarse-grained density and nematic order parameter fields are then defined as 
\begin{align}
    \rho(\mathbf{x})&= \frac{k}{\pi \max_j [r_j^2]} \, ,\\
    q(\mathbf{x})&= \dfrac{\sum_{j=1}^k w_j e^{2i \phi_j}}{\sum_{j=1}^k w_j} \,,
\end{align}
where $\max_j$ denotes maximum among the $k$-neighbors and
\begin{align}
w_j &= \exp\left[{-\dfrac{r_j^2}{\tfrac{1}{k}\sum_{j=1}^k r_j^2}} \right]\,.
\end{align}
The complex coarse-grained order parameter field $q(\mathbf{x})$ can also be written as $q(\mathbf{x})=S(\mathbf{x})e^{2i\theta(\mathbf{x})}$, where $S(\mathbf{x})=|q(\mathbf{x})|$ is the magnitude of the order parameter field and $\theta(\mathbf{x})$ the director angle field. In our analysis below we use $k=10$ and have verified that our results are robust to variations of $k$.

\begin{figure}
    \centering
    \includegraphics[width=0.8\linewidth]{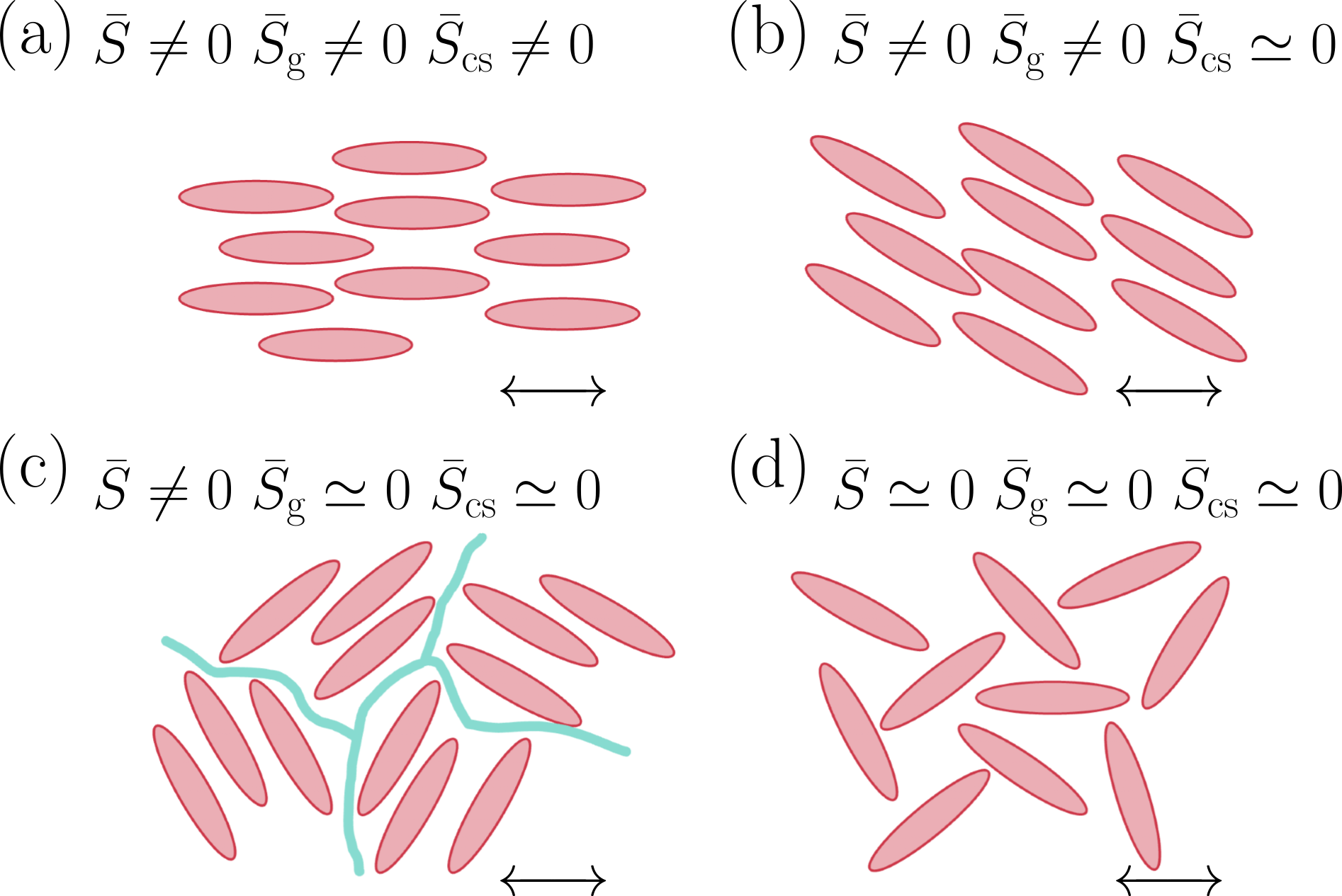}
    \caption{\rv{Cartoons to illustrate the distinction between average local cell-cell order ($\bar{S}$), global cell-cell order ($\bar{S}_\mathrm{g}$), and cell-substrate order ($\bar{S}_\mathrm{cs}$). The direction of nematic order of the substrate is taken as the $x$-axis (double arrows).}}
    \label{fig:order-cartoons}
\end{figure}

To quantify nematic order we introduce three average quantities:
\begin{align}
    \bar{S} &= {\langle |q\rho|\rangle}_{\mathbf{x}}/\langle\rho\rangle_{\mathbf{x}}\,,\\
    \bar{S}_\mathrm{g} &= \sqrt{|\langle q\rho\rangle_{\mathbf{x}}|^2}/\langle\rho\rangle_{\mathbf{x}}\,,\\
    \bar{S}_\mathrm{cs} &= \langle \mathrm{Re} \, q\rho \rangle_{\mathbf{x}}/\langle\rho\rangle_{\mathbf{x}}\,,
\end{align}
where the brackets denote a spatial average over the entire system. The quantities are then averaged over ensembles.

Here $\bar{S}$ measures average local cell-cell nematic alignment and is finite in a state with large randomly oriented nematic domains with an intermediate size that is much larger than the cell spacing but smaller  than the system size.
It will be referred to in the following as the local order parameter.  
$\bar{S}_{\mathrm{g}}$ quantifies global cell-cell order and $\bar{S}_\mathrm{cs}$ captures global alignment with the substrate's direction of broken symmetry, chosen by convention as the $x$-axis. The cell-substrate order, $\bar{S}_\mathrm{cs}$, ranges from $-1$ for monolayers aligned with the y-axis to $+1$ for monolayers aligned with the x-axis. Note that $\bar{S}_\mathrm{cs} < \bar{S}_{\mathrm{g}} < \bar{S}$, i.e.
 local order is a requisite for global order, and global order is a prerequisite for cell-substrate order.
The distinction between the three order parameters is illustrated in the cartoons shown in Fig.~\ref{fig:order-cartoons}.
A monolayer globally aligned to the substrate has high values of $\bar{S}_{\mathrm{cs}}$, $\bar{S}_{\mathrm{g}}$ and $\bar{S}$ (see Fig.~\ref{fig:order-cartoons}(a)). A monolayer that is globally ordered but not aligned with the substrate would have finite $\bar{S}_{\mathrm{g}}$ and $\bar{S}$, but small $\bar{S}_{\mathrm{cs}}$ (see Fig.~\ref{fig:order-cartoons}(b)).
A monolayer with aligned but randomly oriented domains has finite local order $\bar{S}$, but vanishing global and cell-substrate order in the limit of large system size ($\bar{S}_{\mathrm{g}}\simeq 0$ and $\bar{S}_{\mathrm{cs}}\simeq 0$, see Fig.~\ref{fig:order-cartoons}(c)).
Finally, a monolayer that is disordered on the cell scale has $\bar{S} \approx 0$ and thus also vanishing global and cell-substrate order (see Fig.~\ref{fig:order-cartoons}(d)).
Global order clearly also implies local order, but not the opposite. Similarly, a small value of global order necessitates low cell-substrate order, but not vice versa.

The cell-substrate order shows a strong  dependence on the density, as can be seen in Fig.~\ref{fig:experiment_fig_1}(d). Below $\SI{300}{cells/mm^2}$, there is little-to-no cell-substrate order in the monolayer. Order, however, develops sharply at higher density and plateaus around $\SI{550}{cells/mm^2}$ at $\bar{S}_\mathrm{cs} \approx 0.7$, where the monolayer jams orientationally. We see a significant dependence of the cell-substrate order at jamming on the initial seeding density, namely smaller seeding densities lead to more aligned monolayers; see Fig.~\ref{fig:experiment_fig_1}(e).

The orientation of individual cells is affected by mechanical stress exerted by neighboring cells, and hence, by the density of cells. We observe that the cells' orientation tends to be orthogonal to density gradients (see Fig.~\ref{fig:experiment_fig_1}(c)). In other words, cells tend to align parallel to the interface separating high and low density regions. The same behavior has been observed before in proliferating bacteria~\cite{DellArciprete2018ANematic} and is consistent with what is expected in extensile active fluids~\cite{Blow2014BiphasicNematicsb}. In the following we use the above observations to formulate a continuum nemato-hydrodynamic theory.

\section{A Proliferating Nematic Capable of Jamming}
\label{sec:model}

To capture the interplay between cell growth and orientation, we consider a minimal continuum model where the steady-state nematic texture minimizes the Landau-de Gennes free energy, given by 
\begin{align}
    \mathcal{F}=\int \mathrm{d}^2 r \left[f_\mathrm{LdG}(\mathbf{Q}) + f_\chi (\mathbf{Q})+f_{\Pi}(\mathbf{Q})\right]\,,
    \label{eq:FLdG}
\end{align}
with free energy densities 
\begin{align}
        \label{eq:F_LG}
        f_\mathrm{LdG}(\mathbf{Q}) &= f_\mathrm{LdG}^0(\mathbf{Q}) + \frac{K \rho}{2 \rho_\mathrm{IN}^{}} |\mathbf{\nabla Q}|^2\,, \\
        \label{eq:F_LG0}
        f_\mathrm{LdG}^0(\mathbf{Q}) &= \frac{a(\rho)}{2} |\mathbf{Q}^2| + \frac{b(\rho)}{4} |\mathbf{Q}^2|^2]\,,\\
        \label{eq:F_chi}
        f_{\chi}(\mathbf{Q}) &= \tilde\chi Q_{ij} (\partial_i \rho)(\partial_j \rho) \,,\\
        \label{eq:F_pi}
        f_{\Pi}(\mathbf{Q}) &= -\tilde\Pi\rho P_{ij} Q_{ij} \,.
\end{align}
The real order parameter field $Q_{ij}(\mathbf{x})$ is a symmetric traceless tensor related to the complex order parameter $q(\mathbf{x})$ by $q/2=Q_{xx}+iQ_{xy}$, with $Q_{yy}=-Q_{xx}$ and $Q_{yx}=Q_{xy}$. It can also be written in the familiar form $Q_{ij} = S(n_i n_j-\delta_{ij}/2)$, where $S$ measures the degree of nematic order and $\mathbf{n}=\left(\cos\theta,\sin\theta\right)$ is the director field that identifies the direction of broken orientational symmetry, with $\mathbf{n}=-\mathbf{n}$ and $|{\mathbf{n}}|=1$. 
The first two terms in Eq.~\eqref{eq:F_LG} control the equilibrium state in the absence of substrate coupling, with $a(\rho)=a_0(1-\rho/\rho_\mathrm{IN}^{})$ and $b(\rho)=2a_0(1+\rho/\rho_\mathrm{IN}^{})$. The third term denotes nematic elasticity in a one constant approximation, which we scale with density, as is seen in simulations of hard rods~\cite{Revignas2023OnCrystals}. When $\tilde\Pi=0$ the steady state is controlled by the isotropic--nematic transition density $\rho_\mathrm{IN}^{}$. For $\rho<\rho_\mathrm{IN}^{}$ the cell layer is isotropic, with $S=0$. For $\rho>\rho_\mathrm{IN}^{}$ the cell layer is in a nematic state, with $S=\sqrt{(\rho-\rho_\mathrm{IN}^{})/(\rho+\rho_\mathrm{IN}^{})}$,
which saturates to $1$ for $\rho\gg\rho_\mathrm{IN}^{}$.
In the absence of alignment terms ($\chi = \tilde{\Pi} = 0$), the direction of broken symmetry is chosen spontaneously. In the following all densities are measured in units of $\rho_\mathrm{IN}^{}$, unless otherwise indicated.

The free energy density $f_\chi$ couples orientation to density gradients, favoring  alignment of the director normal to density gradients for $\tilde\chi>0$ (as seen in Fig.~\ref{fig:experiment_fig_1}(c)). We define a non-dimensional parameter $\chi=\tilde\chi \rho_\mathrm{IN}^2 /K$ to refer to this coupling strength.
Finally, $f_\Pi$ describes the interaction with the substrate modeled as an external field $P_{ij}$ that aligns cells with the direction of nematic order in the substrate.
In the following we consider a uniform substrate anisotropy aligned, by convention, with the $x$-axis: $P_{xx}=-P_{yy}=1$ and $P_{xy}=P_{yx}=0$. The strength of alignment is controlled by the energy scale $\tilde\Pi$. Collective sensing of the substrate is incorporated via the factor of cell density in $f_\Pi$. \rv{We note that a similar extra-cellular guiding term with density and order dependence was proposed earlier to explain density dependent isotropic--nematic transition in human Melanocytes~\cite{Kemkemer2000NematicCells}}.

\rv{The external field term is a phenomenological stand-in for the many ways environmental anisotropic forces affect cells. Examples include vascular endothelial cells aligned by fluid shear flow~\cite{Resnick2003FluidWorse}, fibroblasts that align with the extracellular matrix (ECM) they deposit~\cite{Li2017OnFibroblasts, Duclos2014PerfectCells, Adar2024Environment-StoredRemodeling, Guillamat2025GuidanceSurfaces}, non-motile bacteria under spatial confinement~\cite{Volfson2008BiomechanicalPopulations, You2021Confinement-inducedColonies}, confined amniserosa cells in drosophila embryo~\cite{Ray2024ConfinementEmbryogenesis}, actin fibers in \emph{Hydra} guided morphogen gradients~\cite{Maroudas-Sacks2020TopologicalMorphogenesis}, and various elongated cells guided by micro- or nano-grooves on surfaces~\cite{Lacroix2024EmergenceGuidance}.}

To describe the ordering dynamics \rv{during cell growth, we augment a well established continuum model of active nematic liquid crystals ~\cite{Doostmohammadi2018ActiveNematics} to include cell proliferation and incorporate  the density dependence of various alignment rates.} We assume that cells are seeded at a density $\langle \rho(t = 0) \rangle = \rho_\mathrm{seed}$ and then proliferate, with the density growing logistically up to a characteristic carrying capacity $\rho_{c}(S)$,  (defined as the number of cells that can attach to a given substrate area), which depends on the local nematic order. This dependence incorporates the fact that an aligned monolayer can accommodate a larger carrying capacity. The biological motivation for this is the reduced density seen in defect cores which are regions of vanishing nematic order. The nematic order parameter in turn evolves with relaxational dynamics controlled by the free energy $\mathcal{F}$, while getting advected, rotated, and aligned by cellular flows with velocity $\mathbf{v}$. The coupled dynamics of density, nematic order \rv{ and flow} is then given by
\begin{align}
    &\partial_t \rho + \bm\nabla\cdot(\rho\mathbf{v}) = \frac{\rho}{\tau_g} \left[ 1 - \frac{\rho}{\rho_\mathrm{C}(S)}\right] \,,
    \label{eq:rho_eq}\\
    &\mathcal{D}_t \mathbf{Q} = \lambda(\rho) \bm E - \frac{1}{\gamma(\rho)} \left[\tfrac{\delta \mathcal{F}}{\delta\mathbf{Q}}\right]^T
    + \Delta(\rho)\,\mathbf{N}\,, \label{eq:Q-eq}\\
   &\rv{\rho\zeta_{ij}v_j} = \rv{\partial_j\sigma^a_{ij}-\partial_ip\,.}
   \label{eq:Stokes}
\end{align}
Here $\tau_g$ is the proliferation time at low density \rv{which we use to nondimensionalize all timescales in the system} and $\mathcal{D}_t\mathbf{Q} = \partial_t \mathbf{Q} + \nabla_k (v_k \mathbf{Q}) + \mathbf{Q}\cdot\bm\Omega - \bm\Omega \cdot\mathbf{Q}$, with $\Omega_{ij} = (\partial_j v_i - \partial_i v_j)/2$ and 
$E_{ij} = (\partial_j v_i + \partial_i v_j - \delta_{ij}\partial_k v_k)/2$. Here
$[\cdot]^T$ denotes the traceless part.
\rv{In experiments, it is observed that cell orientation dynamics arrests upon jamming at high density. We therefore include a density dependence of the nematic ``mobility'' and flow alignment rate so that both vanish at the jamming density $\rho_\mathrm{J}(S)$:}
\begin{align} 
    \label{eq:gamma}
    \frac{1}{\gamma(\rho)} &= \frac{1}{\tilde\gamma_0}~\Theta{\left[\rho_\mathrm{J}(S)-\rho\right] }\,,\\
    \label{eq:lambda}
    \lambda(\rho) &= \lambda_0~ H{\left[\rho_\mathrm{J}(S)-\rho\right] } \,,
\end{align}
where $\Theta(x)= x H(x)$, with $H(x)$ the Heaviside step function. The jamming density $\rho_\mathrm{J}$ is assumed to depend on the degree of nematic order,
\begin{align} \label{eq:rhoJ}
    \rho_\mathrm{J}(S) = \rho_\mathrm{J}(0) + [\rho_\mathrm{J}(1)-\rho_\mathrm{J}(0)] S^2\,,
\end{align}
where $\rho_\mathrm{J}(0)$ and $\rho_\mathrm{J}(1)$ are the jamming densities in the isotropic and nematic state, respectively.
We set $\rho_\mathrm{J}(1) > \rho_\mathrm{J}(0)$ because aligned cells can pack better, and thus, jam at a higher density.
Estimates of the various densities in the system are provided in Appendix~\ref{App:density_estimate}.

\rvv{The dependence of the nematic relaxation rate on jamming density, which is itself controlled by the scalar order parameter, introduces a feedback between density and order, which is inherently non-equilibrium in nature and cannot be captured by equilibrium Landau-like theories. We show in the next section how this feedback encodes a dependence on initial conditions.}

In general the density $\rho_\mathrm{J}$ where orientational order freezes and the density $\rho_\mathrm{C}$ where cells stop dividing (the carrying capacity) can be different. For simplicity, we choose their ratio  to be constant, with 
\begin{align}
    \rho_\mathrm{C}(S) = r_\mathrm{J} \rho_\mathrm{J}(S)\,.
\end{align}
The ratio $r_\mathrm{J}$ will be an important parameter of the model. This choice is motivated by the form of the equation of state, where pressure is chosen to increase exponentially as the system reaches jamming~\cite{Volfson2008BiomechanicalPopulations},
\begin{align}
    p(\rho, S) = B e^{\rho/\rho_\mathrm{J}(S)}\,.
    \label{eq:pressure}
\end{align}
Here, $B$ plays the role of a bulk modulus, akin to an inverse compressibility.
As the system reaches carrying capacity $\rho = \rho_\mathrm{J}(S)$, pressure becomes uniform, implying that density inhomogeneities can exist post jamming without creating flows.

Finally, the $Q$-dynamics includes a non-conserved Gaussian white noise $N_{ij}(\mathbf{x},t)$ with zero mean, variance $\Delta(\rho) = \eta_Q \rho/\gamma(\rho)$ and correlations
\begin{align}
    \langle{N}_{ij} (\mathbf{x}, t) {N}_{kl}(\mathbf{x}', t') \rangle &= J_{ijkl}\delta(\mathbf{x}-\mathbf{x}') \delta(t-t')  \,,
\end{align}
with $J_{ijkl} = \left(\delta_{ik}\delta_{jl} + \delta_{il}\delta_{jk}-\delta_{ij}\delta_{kl}\right)/2$.

\rv{The force balance equation \eqref{eq:Stokes} contains an anisotropic friction $\zeta_{ij}=\zeta_0(\delta_{ij}-\epsilon P_{ij})$, where $P_{ij}$ encodes the substrate orientation and $\epsilon$ is a dimensionless parameter that controls the friction anisotropy.}

The first term of the right hand side of Eq.~\eqref{eq:Stokes} is the gradient of the active stress with deviatoric component
\begin{align} \label{eq:sigma-a}
    \sigma_{ij}^a = \alpha \Theta[\rho_\mathrm{J}(S)-\rho] Q_{ij}\,,
\end{align}
where $\alpha<0$ corresponds to extensile activity and the pre-factor is chosen to make the active stress vanish as the cells jam.
\rv{Note that the active stress also vanishes for small $\rho$ because the magnitude of $Q$ vanishes below the isotropic-nematic transition density $\rho_\mathrm{IN}$.}

\begin{figure*}[t]
    \centering
    \includegraphics[scale=0.75]{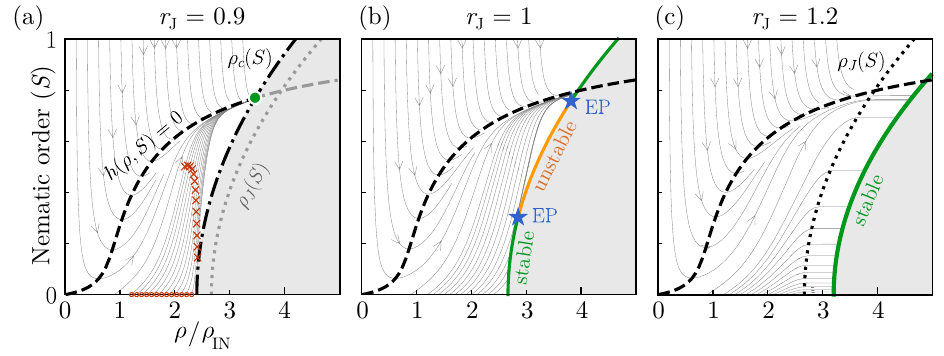}
    \caption{
    Phase portraits of the ODE model showing the dynamics as light gray trajectories. The black dashed line in each plot is the $S$-nullcline obtained from the solution of $h(\rho,S)=0$. The dotted and dash-dotted lines show the jamming density $\rho_\mathrm{J}(S)$ and carrying capacity $\rho_\mathrm{C}(S)$, respectively.
    (a)~Phase portrait for $r_\mathrm{J}=0.9$ with a single stable fixed point (green dot; $ h(\rho_\mathrm{C}, S)=0 $). The red crosses show the state at $T=2\tau_g$ for simulations started from $S = 0$ and $\rho$ uniformly spaced between $1.2$ and $2.5$ (red dots) illustrating that the dynamics slows down as the seeding density is increased.
    (b)~Phase portrait for $r_\mathrm{J}=1$ where the jamming and carrying capacity lines coincide, forming a line of fixed points (solid line).
    Green/orange segments correspond to stable/unstable fixed points. Blue stars indicate exceptional points where the two eigenvalues coalesce.
    (c)~Phase portrait for $r_\mathrm{J}=1.2$ showing the line of fixed points (solid green line) which is always stable.
    Parameters: $\Pi = 0.02$, $\gamma_0 = 0.67$.
    }
    \label{fig:phase_space_sketch}
\end{figure*}

Fibroblast colonies exhibit various active processes that may be relevant on different time and length scales. Individual cells crawl on the substrate at speeds of around \SI{10}{\micro m/h}~\cite{Luo2023Molecular-scaleMonolayers} and reverse their direction of motion roughly every hour~\cite{Duclos2014PerfectCells}. On time scales shorter than the reversal time, cells move by generating contractile force dipoles. At longer times the back-and-forth motion along the cell's long axis effectively yields extensile stresses. \rv{For elongated cells that divide along their long axis, growth and division are a source of extensile activity~\cite{Ranft2010FluidizationApoptosis}}. The cell's doubling time is around 50--\SI{60}{h}~\cite{Luo2023Molecular-scaleMonolayers}. In this study, we focus on the two dominant long-time sources of anisotropic active stresses in the monolayer, both extensile in nature: cell growth and back-and-forth motion of cells along their long axis.
Notably, as we shall see below, these anisotropic active stresses are not essential for the collective alignment of cells on an anisotropic substrate. Rather it is the isotropic proliferation-induced pressure gradient that drives collective alignment.

\section{A lower seeding density yields a more ordered state}
\label{sec:homogeneous}

\rv{The dependence of the structure of the final jammed state on seeding density can be captured through a simple model that neglects all spatial variations. It simply originates from the} interplay between growth, jamming, and alignment to the substrate. 

\rv{Neglecting all gradients,} the coupled dynamics of density, nematic \rv{order parameter and director angle is given by}

\begin{align}
\label{eq:rho}
    \frac{d\rho}{dt} &= \rho\left(1-\frac{\rho}{r_\mathrm{J}\rho_\mathrm{J}}\right)\,,\\
    \label{eq:S}
    \frac{dS}{dt} &= -\frac{1}{\gamma_0} \Theta{(\rho_\mathrm{J}-\rho)}~ h(\rho, S) \,,\\
    \label{eq:theta}
    \frac{d\theta}{dt} &= -\frac{1}{\gamma_0 S}  \Theta{(\rho_\mathrm{J}-\rho) \Pi\rho} \sin{2\theta} \,,
\end{align}
\rv{where $h(\rho, S) = S(1-\rho+ (1+\rho) S^2) - 2\Pi\rho$. We have} rescaled times with $\tau_g$, densities with $\rho_\mathrm{IN}^{}$, and energies with $a_0$. \rv{The equations now contain three dimensionless parameters: $\Pi=\tilde\Pi/a_0$, $\gamma_0=\tilde\gamma_0/(a_0\tau_g)=\tau_Q/\tau_g$, where $\tau_Q=\tilde\gamma_0/a_0$ is the nematic relaxation time,} and $r_\mathrm{J}$  is the ratio of the carrying capacity to the density at which orientations jam. These equations are solved with initial condition $\rho(t=0)=\rho_\mathrm{seed}$, $S(t=0)=0$ and randomly chosen $\theta(t=0)$. In the homogeneous case, once the angle variable aligns due to the external field, the average local order parameter ($\bar{S}$), the global order parameter ($\bar{S}_\mathrm{g}$), and the cell-substrate order parameter ($\bar{S}_\mathrm{cs}$) all coincide and will be simply denoted by $S$.

The steady states of these ODEs and their stability are summarized in Fig. \ref{fig:phase_space_sketch}. The angle $\theta$ aligns with the nematic direction of the substrate and the equations can be simplified to just examine the \rv{coupled dynamics of $S$ and $\rho$. There are two nullclines (defined by $\frac{d\rho}{dt}=0$ and $\frac{dS}{dt}=0$)} in the $S$-$\rho$ plane. We find it instructive to divide our analysis into three cases based on the value of $r_\mathrm{J}$.

When $r_\mathrm{J}<1$, or the carrying capacity of the monolayer is smaller than the jamming density, the only steady state is a stable fixed point (green dot in Fig.~\ref{fig:phase_space_sketch}(a)) given by: $h(\rho_\mathrm{C}(S), S)=0$. In this case the late-time behavior of the system is independent of the seeding density. \rv{However, the dynamics of approach to the fixed point depends strongly on seeding density.}  When the system is initialized at high seeding density the dynamics significantly slows down near $\rho_\mathrm{J}(S)$. 
On the other hand, trajectories starting at low seeding density rapidly approach the fixed point, since they are far away from $\rho_\mathrm{J}$. To illustrate this, we indicate the state of the system  at $T=2\tau_g$ by gray crosses in Fig.~\ref{fig:phase_space_flows}(a), showing that the \rv{rate of} build-up of order is inversely proportional to the seeding density.

This behavior has dramatic consequences at $r_\mathrm{J}=1$, where two nullclines coincide and the jammed state $\rho=\rho_\mathrm{J}(S)$ is a steady state for any value of $S$. This condition identifies a line of fixed points shown as a solid line in Fig.~\ref{fig:phase_space_sketch}. The stability of the fixed points determines the behavior of the trajectories. Stable (green) and unstable (orange) segments are separated by Exceptional Points (EP). In Appendix~\ref{App:linear_stability} we derive conditions for the presence and positions of the pair of EPs and create a bifurcation diagram (see Fig.~\ref{fig:phase_space_flows}) that clearly shows that a part of the phase-space becomes inaccessible to the system due to the emergence of an unstable segment along the line of fixed points.

\begin{figure*}
    \centering
    \includegraphics[width=0.9\linewidth]{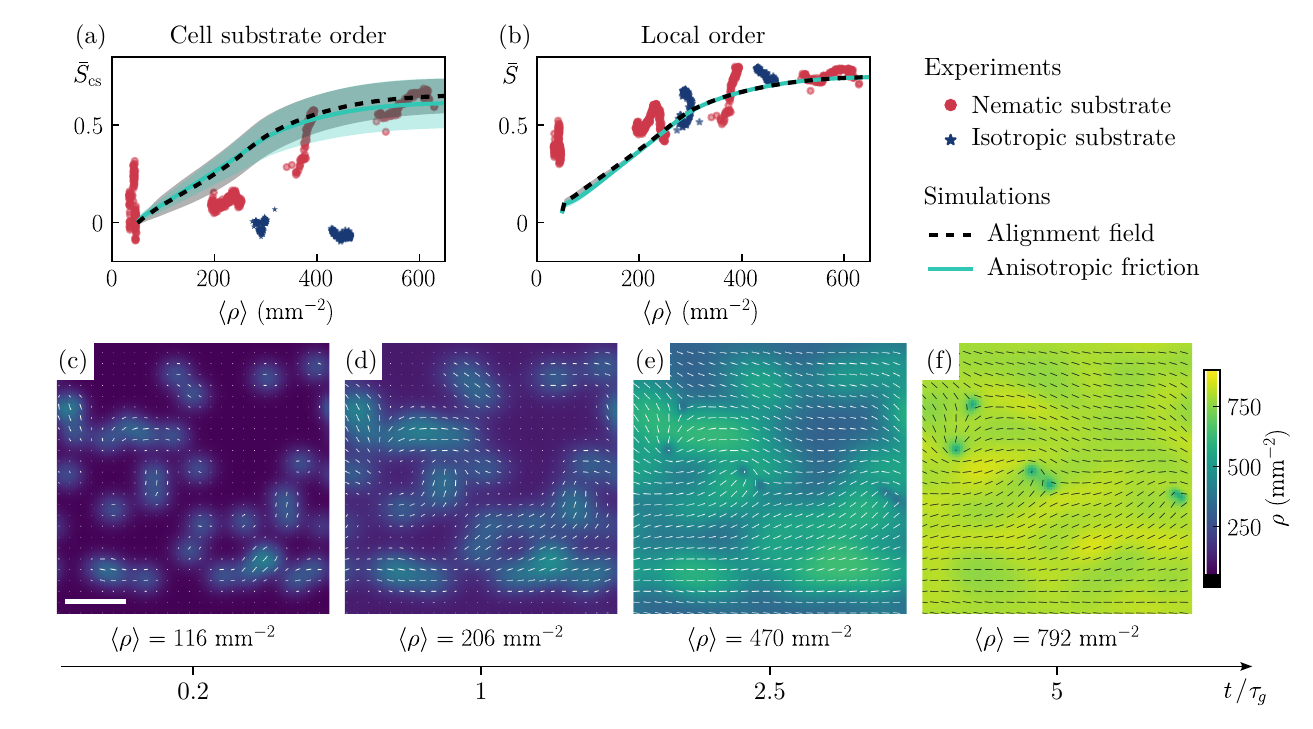}
    \caption{
    Emergence of order in numerical simulations of a proliferating nematic. 
    (a)~Cell-substrate order as a function of mean density.
    (b)~Local order parameter as a function of mean density. 
    Black and cyan lines show results for the model variant with an external aligning field (average of 10 runs) and anisotropic friction (average of 20 runs), respectively.
    Red disks and blue stars show experimental data for cells growing on nematic vs isotropic substrates respectively.
    Simulations for external field and asymmetric friction are for $n=(10, 20)$ runs per $\rho_\mathrm{seed}$ respectively.
    (c--f) Snapshots of a simulation with asymmetric friction starting at $\rho_\mathrm{seed}= \SI{100}{mm^{-2}}$ showing the local nematic director (dashes) and the local density (color-map). Initially (c) cells are clustered and locally aligned within the cluster, but the clusters are randomly oriented. As cells divide, and the mean densities increases clusters grow, merge, and rotate to align with the substrate (d). As density increases everywhere, significant local order develops in the entire monolayer (e). At the highest density (f) the orientations jam. Comparing (e) and (f) one can see that jamming prevents annihilation of defects pairs. Scale bar: $\SI{2}{mm}$.
    For details about the simulation parameters refer to Sec.~\ref{subsec:numerical_sims}. 
   }
    \label{fig:sim_results_both}
\end{figure*}

Finally we examine the case $r_\mathrm{J}>1$, where the carrying capacity of the monolayer is larger than the jamming density. In this case $\rho=r_\mathrm{J} \rho_\mathrm{J}$ is a line of fixed points due to the Heaviside function in Eq.~\eqref{eq:S}. This line of fixed points is always stable and supports the observation that the seeding density inversely determines the degree of final order. Trajectories that start equally spaced from the $x$-axis cluster towards small and large $S$, and become diluted for intermediate $S$ even though the entire line is stable. This behavior can be understood as a ``ghost'' of the segment of unstable fixed points found for $r_\mathrm{J}=1$ at intermediate $S$. The dynamics in the regime $r_\mathrm{J}>1$ robust for small noise in both variables and our full spatiotemporal simulations, described in the next section, are carried out in this regime.
The results of the numerical solution of Eqs.~\eqref{eq:S}-\eqref{eq:rho} with $r_\mathrm{J}=1.5$ are shown in Fig.~\ref{fig:experiment_fig_1}(a,b). These show that the model successfully captures the dependence of order at jamming on the seeding density.

We can intuitively understand the observed dynamics as the result of two competitions: a competition of timescales and a competition of alignment mechanisms. The two competing timescales are the rate of growth of density, which drives the density towards its jamming value $\rho_\mathrm{J}(S)$, and the rate $h(\rho,S)$ at which steric effects and the external field drive nematic order.
To reproduce the experimental observations, ordering must be sufficiently fast for cells to align before jamming occurs, but not too fast such that high seeding density results in jamming before an ordered state has been reached. 
The competition between steric alignment and alignment driven by the external field controls the behavior of the ordering rate $h(\rho,S)$. 
For small values of $\Pi$ the external field selects the direction of broken symmetry, but the onset of order is dominated by steric effects and the isotropic to nematic transition density $\rho_\mathrm{IN}^{}$ determines the threshold for the onset of order. For large $\Pi$ alignment is controlled by the external field. In this case the system reaches an ordered state for all seeding densities. For intermediate values of $\Pi$ the ordering rate $h(\rho, S)$ is fast enough to align before jamming. In this regime smaller seeding densities reach more ordered states.

\begin{figure}
    \includegraphics[width=\linewidth]{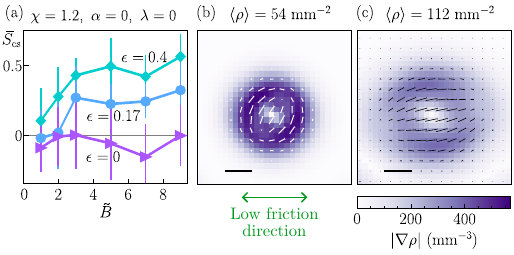}
    \caption{
    Collective alignment in simulations with asymmetric friction.
    (a)~Cell-substrate order at jamming as a function of the non-dimensionalized bulk modulus $\tilde{B} = B a_0 \tau_g/(\zeta_0 K \rho_\mathrm{IN})$ for three different values of friction anisotropy: $\epsilon=0.4$ (cyan diamonds), $\epsilon=0.17$ (blue dots) and $\epsilon=0.0$ (purple triangles). Lines are drawn for visual clarity. Simulations were performed with finite density alignment ($\chi=1.2$), but no flow alignment nor activity ($\lambda=0$, $\alpha=0$), with $\rho_\mathrm{seed}=\SI{50}{mm^{-2}}$.
    (b,c)~Magnitude of density gradients (heat-map) and local nematic director (white segments) for a single cell patch in a simulation with asymmetric friction at two different times, corresponding to two values of mean density.
    At earlier times, the gradients have azimuthal symmetry and the director is oriented azimuthally, with broad regions where it is normal to the $x$ axis (b).
    After doubling of the mean density, the density gradients are stronger in the in $y$-direction, the patch itself is elongated and cells have aligned with the $x$ axis (c). Scale bars:$\SI{500}{\micro m}$.
    }
    \label{fig:sim_scs_params_sweep}
\end{figure}

\section{Spatially extended model}
\label{sec:inhomogeneous}

The minimal set of ODEs described in the previous section explains how the emergence of order depends on the seeding density. It does not, however, distinguish between local and global order nor does it address the question of \emph{how} cells \rv{collectively} align with the substrate. It is  known that inhibiting focal adhesions results in the loss of collective directional sensing~\cite{Luo2023Molecular-scaleMonolayers}.
We therefore hypothesize that cell attachment depends on the cell's orientation relative to the substrate order due to the microscopically anisotropic structure (polymer chains) of the LCE which may affect focal-adhesions.
\rv{We incorporate this in our model as anisotropic traction forces responsible for the anisotropic friction in Eq.~\eqref{eq:Stokes}.
In the simulations with anisotropic friction, we set the ``external field'' $\Pi = 0$. Simulations with isotropic friction and $\Pi > 0$ are presented in Appendix~\ref{app:external-field}. As expected, this model variant also reproduces the experimental findings but doesn't explain the mechanism underlying collective alignment.
}

\rv{To mimic the initial conditions found in experiments after cell seeding, we initialize simulations with randomly oriented cell clusters (or colonies as seen in Videos ($1$--$3$) with descriptions in the Supplementary Material ~\cite{Parmar2025SupplementalSubstrate}); see Appendix~\ref{App:clustering} for details.
Starting from a global average density $\rho_\mathrm{seed} = \SI{50}{cells/mm^{2}}$, we find that both local order and cell-substrate order increase over time as cells proliferate, reproducing the experimental observations; see Fig.~\ref{fig:sim_results_both}(a--b). 
In particular, at intermediate density (${\sim}\SI{200}{cells/mm^{2}}$) there is appreciable local order but little cell-substrate order (see lower middle panels in Videos ($1$--$3$)).
Increasing the seeding density results in reduced cell-substrate alignment in the final, jammed state; see Fig.~\ref{fig:Scs-rho_seed} in Appendix~\ref{app:external-field}. Thus, the the spatially extended model reproduces the experimental findings and the results obtained from the minimal ODE model.
We conclude that anisotropic friction is sufficient for cells to collectively align with the substrate direction.
We next investigate the underlying mechanism.}

\subsection{Anisotropic friction and density gradients drive collective alignment}
\label{subsec:collective_alignment}

\rv{To reveal the key processes driving collective cell alignment in our model, we performed simulations with flow alignment and active deviatoric stresses turned off ($\lambda = 0, \alpha = 0$). 
Remarkably, these simulations still show significant cell substrate order at jamming Fig.~\ref{fig:sim_scs_params_sweep}(b).
Parameter sweeps of the bulk modulus ($B$) show that proliferation-driven stress plays a key role.}

\rv{
In the presence of anisotropic friction, this isotropic stress drives anisotropic flows which are stronger along the low-friction direction.
As a result, density gradients become shallower along this direction compared to the high-friction direction. 
In turn, anisotropic density gradients cause reorientation of the nematic through alignment perpendicular to density gradients. Indeed, setting the density-alignment parameter $\chi = 0$ results in a significant reduction of cell-substrate order, even in the presence of flow alignment and deviatoric active stresses [see Fig.~\ref{fig:gradient_asymmetry}(a,b) in Appendix~\ref{App:collective_align}].}

\rv{
To illustrate the above mechanism in simulations, we looked at a single initially circular cell cluster with initially uniform nematic order at an angle relative to the low-friction direction. Alignment of cells to density gradients rapidly orients the nematic circumferentially around along the cluster edge; see Fig.~\ref{fig:sim_scs_params_sweep}(b).
A snapshot at a later time, Fig.~\ref{fig:sim_scs_params_sweep}(c), clearly shows the anisotropic density gradients which are smoother along the low-friction direction ($x$-axis) and the resulting reorientation of the nematic directors toward that axis.
To test whether this mechanism may explain the emergence of cell-substrate order in experiments, we quantified gradients of cell density. 
In agreement with our theory, we find that gradients perpendicular to the direction of substrate order are steeper; see Fig.~\ref{fig:gradient_asymmetry}(b) in Appendix~\ref{App:collective_align}.
}
\rvv{We must emphasize that initial density gradients are a necessary ingredient for this mechanism. The density heterogeneity in the initial state which gives rise to these gradients is estimated from the experiments (see Appendix~\ref{App:clustering} for details).
In contrast to this initial heterogeneity, dynamical noise added to Eq.~\eqref{eq:Q-eq} does not significantly affect the behavior of the system.
}

\begin{figure}
    \includegraphics[width=\linewidth]{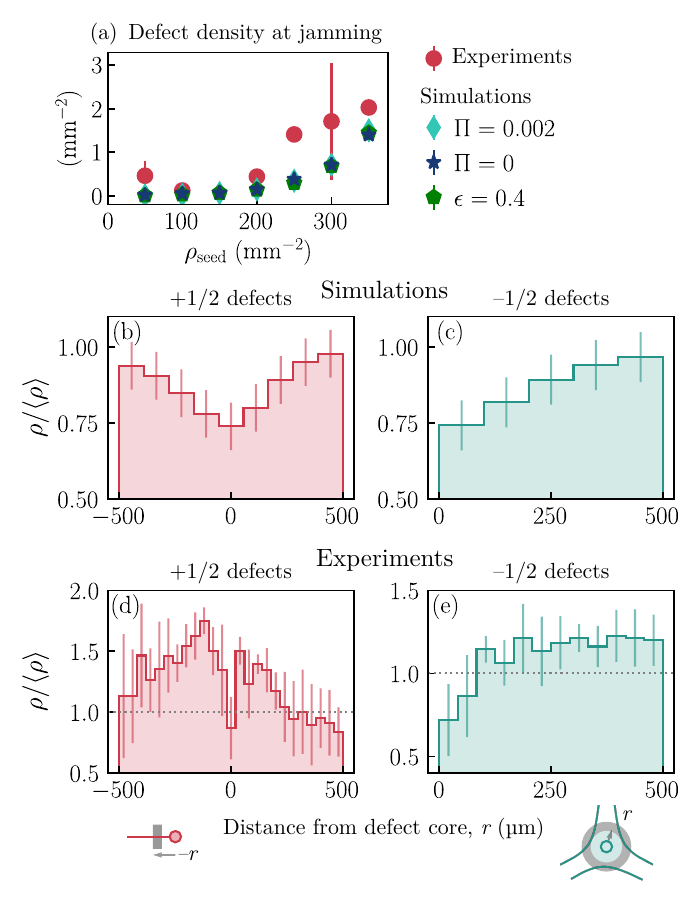}
    \caption{
    Topological defects in simulations and experiments.
    (a)~Defect number density (units of $\unit{mm^{-2}}$) at jamming increases with seeding density in experiments on nematic substrates (red dots) and in simulations with (cyan diamonds) and without (blue stars) external aligning field as well as simulations with (green pentagons) asymmetric friction.
    (b) and  (c)~Histograms of relative cell density at a distance $r$ from the defect core in simulations. Note the low cell density for both  $+ 1/2$ and $-1/2$ defects. Data form $30$ simulations with initial low seeding densities ($50-\SI{150}{mm^{-2}}$) yielding $23$ defect pairs.
    \rv{(d) and (e) ~Experimental density histograms near defects in a jammed monolayer on a nematic substrate. The histograms clearly show lower density at the defect cores. Data from $4$ experiments containing a total of $23$ defects with charge $+1/2$ and $21$ with charge $-1/2$. Error bars show standard deviation.}
    }
    \label{fig:sim_defects}
\end{figure}

\subsection{Topological defect dynamics}
\label{sec:defects}

\rv{Regions of appreciable local cell-cell order consist of aligned domains separated by topological defects (pink and cyan dots in Fig.~\ref{fig:experiment_fig_1}(a,b)).  We find that at intermediate to high mean densities ($500$--$\SI{900}{cells/mm^2}$), the defect cores are  regions of low cell density for both the comet-shaped $+1/2$ (Fig.~\ref{fig:sim_defects}(d)) and the trefoil-shaped $-1/2$ (Fig.~\ref{fig:sim_defects}(e)) defects. This is in contrast to previous studies on different cell types where it was observed that cells accumulate at defect cores (neural progenitor cells~\cite{Zhao2025IntegerMonolayers} and \textit{Myxococcus xanthus} bacteria~\cite{Copenhagen2020TopologicalColonies}) or are even extruded due to the large compressive stress that develop in this region (epithelia~\cite{Saw2017TopologicalExtrusion}). At even higher densities ($\gtrsim \SI{1300}{cells/mm^2}$), however, fibroblasts have been found to form mounds at the core of $+1/2$-defects where the system starts forming a second cell layer~\cite{Sarkar2023CrisscrossSheets}. Details on calculating the density near defect cores are presented in Sec.~\ref{Sec:Img_analysis}.}

We investigated the motion of topological defects in more detail.
Topological defects emerge in cell clusters where ``anchoring'' to the density gradient along the cluster edge induces a net charge of $+1$.
The compensating negative topological charge is initially distributed across the low density regions where $S \approx 0$. As clusters grow into one another, this negative charge localizes in $-1/2$ defects along the domain borders where clusters with different cell orientation meet.
The negative and positive defects eventually annihilate as the monolayer orders (see Videos~($1$--$3$)). Upon orientational jamming, some defect pairs freeze before they can annihilate. \rv{The monolayer then is in an arrested state with jammed topological defects, as previously seen in epithelial cells~\cite{Angelini2011Glass-likeMigration, Garcia2015PhysicsMonolayer, Puliafito2012CollectiveInhibition,Bi2016Motility-drivenTissues,Lawson-Keister2021JammingTissues, Jacques2023AgingRemodeling, Sarkar2023CrisscrossSheets}}.
The number of defect frozen in the jammed state of the monolayer increases with increasing seeding density, as shown in
Fig.~\ref{fig:sim_defects}(a). 
This is expected because starting at higher seeding densities results in a faster approach to jamming, giving defects little time to annihilate.

In the original publication of the experiments~\cite{Luo2023Molecular-scaleMonolayers}, it was noted that defects were seen moving in the direction of their head (signature of extensile activity) when they are oriented in the direction of the substrate's broken symmetry. This was attributed to activity due to proliferation. However, defects were also observed to move in the direction of their tail (signature of contractile activity) when oriented perpendicular to the direction of the substrate's broken symmetry. In our work activity is always extensile and the role of the deviatoric active stress is to stretch the ordered high density clusters into a spindle shape due to extensile defect self-propulsion. At both early and late times in the simulations, we see defects move as expected in extensile systems (see Videos~1--3), in agreement with previous calculations~\cite{Brezin2022SpontaneousDefects}.

Finally, we find that cell density is always reduced at the core of both $+1/2$ and $-1/2$ defects in simulations [Fig.~\ref{fig:sim_defects}(b,c)] -- consistent with the observations from experiments [Fig.~\ref{fig:sim_defects}(d,e)].
In the model, reduced cell density near topological defects arises from our assumption that cells can pack better when aligned [cf.\ Eq.~\eqref{eq:rhoJ}], so that density is higher in regions of high nematic order, while disordered defect cores tend to be depleted of cells. This behavior likely arises from the steric repulsion between the elongated cells and  is similar to what is observed in active nematics made from microtubule-kinesin mixtures, where defect cores are indeed void of microtubules~\cite{Sanchez2012SpontaneousMatter}.

By contrast experiments on bacteria~\cite{Copenhagen2020TopologicalColonies} and epithelial cells~\cite{Saw2017TopologicalExtrusion} have found cell accumulation near $+1/2$ defects.
\rv{The cause for this striking difference might be the very large aspect ratios of fibroblast cells and microtubule bundles compared to those of bacteria and epithelial cells. Further, investigating this phenomenon is an interesting direction for future research.}

\section{Discussion}
\label{sec:discussion}

Motivated by recent experiments~\cite{Luo2023Molecular-scaleMonolayers}, we have developed a hydrodynamic model of a proliferating nematic aligned by external cues and capable of jamming. Our work  captures two important observations seen in the system of hdF-cells growing on a LCE-substrate: the order in the final jammed monolayer depends on seeding density and cell ordering is a \textit{collective} phenomenon. 

We consider two ways of modeling the alignment with the nematic substrate: (\textit{i}) an external field that aligns the director field of the continuum model to the direction of order of the substrate and (\textit{ii}) anisotropic traction forces exerted by cells on the substrate. Importantly, the strength of each of these couplings is proportional to density to capture the collective nature of the alignment. 

A minimal spatially homogeneous model with no activity-driven flows is sufficient to explain the dependence of final order on seeding density. \rv{In short, a lower seeding density gives cells more time to align before jamming, resulting in a more ordered final state, as seen in experiments. This dependence on the dynamic history emphasizes the non-equilibrium nature of the system.} 

\rv{Our analysis of previously published experimental data reveals that cells preferentially align perpendicular to density gradients, i.e.\ along density contour lines. This suggests a mechanism for global cell alignment mediated by anisotropic density gradients, which, in turn, originate from anisotropic substrate tractions on a nematic LCE substrate.
This mechanism is inherently collective in nature, which is consistent with the experimental observation that individual cells do not align with the direction of the LCE substrate \cite{Luo2023Molecular-scaleMonolayers}.
}
\rv{Notably, cell alignment mediated by density gradients and anisotropic friction occurs even in the absence of deviatoric active stresses and flow alignment (Fig.~\ref{fig:sim_scs_params_sweep}(a)) but requires cell proliferation and (initial) density inhomogeneities. 
Therefore, this mechanism is fundamentally different from ones previously considered in the literature which rely on deviatoric active stresses and flow alignment \cite{Thijssen2020ActiveParameter}.}

\rv{Preliminary results suggest that a more heterogeneous initial state can result in a more ordered jammed state. Exploring systematically this type of ``order from disorder'' is an interesting direction for future work. The dependence of the continuum model on the amount of heterogeneity in the initial state may be related to the fact that the continuum model itself misses the discrete nature of cell division and neglects the associated small scale fluctuations that are amplified at later times, especially near jamming.}

\rv{Since the proposed mechanism for cell alignment on an LCE substrate depends on cell  proliferation, we predict that cells should no longer align with the substrate direction when  proliferation is inhibited (e.g.\ by arresting the cell cycle).
The molecular mechanism of the proposed cell-traction anisotropy on an LCE substrate remains an open question. Experiments have shown that suppressing focal adhesions impedes the onset of global alignment~\cite{Luo2023Molecular-scaleMonolayers}, which supports our proposed mechanism and provides a starting point for future experiments.}

\section{Methods}
\label{sec:methods}

\subsection{Image analysis}\label{Sec:Img_analysis}

Images of cell nuclei were segmented using Cellpose3's Cyto3 algorithm \cite{Stringer2025Cellpose3:Segmentation, Stringer2020Cellpose:Segmentation}. The generated regions of interest were measured in Fiji \cite{Schindelin2012Fiji:Analysis}. A k-nearest-neighbor (kNN) method was used to generate smooth fields out of discrete nuclei position and orientation data in python \rv{(see Fig.~\ref{fig:kNN})}. 
The kNN method is used extensively for nonparametric probability density estimation~\cite{Biau2015LecturesMethod} in the data-science community. To our knowledge, it has not been used to construct continuum fields out of discrete data in the biological physics community. The most commonly used method to construct local fields out of discrete quantities is a box based method where boxes of size $l$ inversely proportional to the mean number density of the system are chosen to discretize the system and quantities are calculated as averages within boxes. This methods works well at high density, but it is problematic in systems with large density fluctuations and regions that are `empty' of particles. In contrast, the kNN method performs well even in a system with large number density fluctuations (most active systems display giant number fluctuations). This is because when number density fluctuations are large, it is impossible to choose a single box-size that will work  in both sparser and crowded regions. In contrast, the kNN-method is akin to scaling the box-size inversely with the local number density. The kNN method does, however, have its disadvantages -- mainly that it does not perform well when the fields it is approximating have long-tailed distributions or are at arbitrarily small number densities (for example, our lowest seeding densities, $\sim 30- \SI{40}{mm^{-2}}$). In these cases, we propose an adaptive-kNN method inspired by recent work in information theory~\cite{Zhao2022AnalysisEstimation}. This method adaptively chooses k and performs well over standard kNN method for globally sparse data. In future work we plan to benchmark both methods for single-cell resolution data.

Defects were detected by calculating the local winding number at every pixel of the smooth orientation field (smoothed with a gaussian kernel of size $\SI{100}{\micro m}$). Connected pixels with the same winding number were grouped together as a single defect. Cell density around a $+1/2$ defect was calculated inside a rectangle of length $\SI{500}{\micro m}$ and width $\SI{100}{\micro m}$ centered at the defect core and oriented along the defect direction. Cell density around $-1/2$ defect was calculated inside a circular region of radius $\SI{500}{\micro m}$.

 \begin{table*}[t]
    \centering
    \begin{tabular}{clc}
    \hline
    \toprule
        \textbf{\hspace{2em}Parameter \hspace{2em}} & \textbf{Description} & \textbf{Value} \\
        \midrule
        $\rho_\mathrm{seed}$  & Seeding Density &$\rho_\mathrm{IN}^{}/3$ \\
        $\tilde{\gamma}_0 $  &  Rotational viscosity & $ \tau_g/40a_0$ \\
        $\lambda$ & Flow-alignment parameter & $1$ \\
        $r_\mathrm{J}$ & Ratio of carrying capacity to jamming density & $1.5$ \\
        $B/\zeta_0$ & Ratio of bulk modulus to friction coefficient & $~15 ~{K\rho_\mathrm{IN}^{}}/{(a_0 \tau_g)}~$ \\
        $\alpha/\zeta_0$ & Ratio of activity to friction coefficient 
        & $~-1.0 ~K/(a_0 \tau_g)~$ \\
        $\eta_Q$ & Nematic order noise amplitude
        & $1/{\rho_\mathrm{IN}^{}}$; $10/{\rho_\mathrm{IN}^{}}$ \\
        $\chi$ & Strength of normal alignment to density gradients
        & $1.2 $ \\
        $\epsilon$ & Anisotropic friction coefficient
        & $0.4$ \\
        $\Pi$ & External field strength
        & $0.002$ \\
        \bottomrule
    \end{tabular}
    \caption{Values of various parameters used for simulation of the full spatially extended model.
    }
    \label{tab:sim_params}
\end{table*}

 \begin{table*}[t]
    \centering
    \begin{tabular}{clc}
        \toprule
        \textbf{\hspace{2em}Density\hspace{2em}} & \textbf{Description} & \textbf{Value} \\
        \midrule
        $\rho_\mathrm{seed}$ & Seeding density $\rho(t=0) = \rho_\mathrm{seed}$ & $50-350~\unit{mm^{-2}}$ \\
        $\rho_\mathrm{IN}^{}$ & Isotropic to nematic transition density & $150~\unit{mm^{-2}}$ \\
        $\rho_\mathrm{J}(S = 0)$ & Arrest of cell reorientation in a disordered monolayer & $400~\unit{mm^{-2}}$ \\
        $\rho_\mathrm{J}(S = 1)$ & Arrest of cell reorientation in an ordered monolayer & $700~\unit{mm^{-2}}$ \\
        $\rho_\mathrm{C}(S)$ & Carrying capacity of the monolayer
        & $r_\mathrm{J} \rho_\mathrm{J} (S)$ \\
        \bottomrule
    \end{tabular}
    \caption{Definitions and values of various densities.
    }
    \label{tab:various_densities}
\end{table*}

\subsection{Numerical simulations}\label{subsec:numerical_sims}

All numerical simulations were performed in Python. Units for space, time, and density were chosen as $\tau_g{\;\simeq\;} \SI{35}{hr}$~\cite{Luo2023Molecular-scaleMonolayers}, $\sqrt{K/a_0} {\;\sim\;} \ell_\mathrm{cell} {\;\simeq\;} \SI{100}{\micro m}$, and $\rho_\mathrm{IN}^{}{\;=\;}\SI{150}{cells/mm^{2}}$. The homogeneous model was integrated with a Backward Differentiation method (BDF) from the SciPy library with $n=8{\times}10^3$ timesteps for a total time $T=40 \tau_g$. For the spatially extended differential equations, we developed a custom Python code with pseudo-spectral solver made available on Github~\cite{Parmar2025Https://github.com/toshi-physics/pyPSS-proliferating-nematic}. All PDE simulations were performed with $n=8{\times}10^4$ timesteps for a total time $T=8\tau_g$. This time was found sufficient for reaching the carrying capacity for all starting densities considered. Unless otherwise specified, the PDE numerical simulations used the parameters given in Table \ref{tab:sim_params}.

Grid spacing, $dx=dy=\sqrt{K/a_0}$, is always smaller than the nematic correlation length for all densities $\rho>\rho_\mathrm{IN}^{}$, $\ell_Q  = \sqrt{K \rho/(a_0 (\rho-\rho_\mathrm{IN}^{}))}$, since $\ell_Q$ is bounded below by $\sqrt{K/a_0}$. Simulation box with periodic-boundary conditions has length $L_x=L_y=100\sqrt{K/a_0}$, which corresponds to $10{\times}\SI{10}{mm^2}$. The field of view in the experiments is $\sim 6{\times} \SI{3}{mm^2}$.
Defects and density were quantified by the same method used in the analysis of experimental data.

\section*{acknowledgement}
We thank Paarth Gulati for helpful discussions. The theoretical modeling by T.P.\ and M.C.M.\ was supported by the National Science Foundation award DMR-2041459. The analysis of imaging data  was additionally supported by the National Science Foundation award OAC-2411043. F.B.\ acknowledges support by the Gordon and Betty Moore Foundation post-doctoral fellowship (grant \#2919).

\appendix

\section{Parameter estimates from experiments} \label{App:density_estimate}

The various densities entering our model are summarized in Table \ref{tab:various_densities}. Here we describe how we have estimated the range of values for these densities.

The density $\rho_\mathrm{IN}^{}$ can be estimated assuming the cells behave like ellipsoidal particles of length ${\sim} \SI{100}{\micro m}$ and aspect ratio of about $6$.  Fibroblasts, however, move back and forth along their long axis~\cite{Duclos2014PerfectCells}, which could increase their effective aspect ratio to 12--15.  The continuous isotropic to nematic transition for hard ellipsoids in two dimensions has been studied numerically for aspect ratio ranging from 2.4 to 9~\cite{Baron2023AnisotropicFields}, where the IN transition happens at packing fractions between 0.85 to 0.45 respectively. We estimate that these cells with higher effective aspect ratio than 9 transition at a packing fraction of about $0.25$, which gives  $\rho_\mathrm{IN}^{}\sim \SI{150}{cells/mm^2}$. For uniform distribution, this density corresponds to 4--5 cells within a circle of radius \SI{100}{\micro m}. Our results do not change qualitatively when changing this value between 100--\SI{175}{mm^{-2}}. The experiments reported in Ref.~\cite{Luo2023Molecular-scaleMonolayers} give $\rho_\mathrm{IN}^{} \sim \SI{320}{mm^{-2}}$. Previous experiments on human Melanocytes have estimated $\rho_\mathrm{IN}^{}\sim\SI{110}{mm^{-2}}$~\cite{Kemkemer2000NematicCells}. 

The various jamming densities can be estimated from the experiments of Ref.~\cite{Luo2023Molecular-scaleMonolayers} as $\rho_\mathrm{J}(0)\sim 300{-}\SI{400}{mm^{-2}}$, and $\rho_\mathrm{J}(1)\sim \SI{600}{mm^{-2}}$. Alternatively, one can estimate $\rho_\mathrm{J}(1)$ by noting that a nematic to solid transition is observed in 2D simulations of hard ellipsoids above a packing fraction of about $0.85$~\cite{Bautista-Carbajal2014PhaseEllipses} for aspect ratios near 5. Since our cells are considerably longer, we estimate this number to be about $\rho_\mathrm{J}(1)\simeq 700\,\unit{mm^{-2}}$.  
Using $\rho_\mathrm{J}(0)\simeq 400\,\unit{mm^{-2}}$, This gives $\rho_\mathrm{J}(0)/\rho_\mathrm{IN}^{} = 2.4$ and $\rho_\mathrm{J}(1)/\rho_\mathrm{IN}^{}=3.47$. We use these parameters in all numerical and simulation studies presented.

\section{Stability analysis for the ODE model}
\label{App:linear_stability}

\begin{figure*}
    \centering
    \includegraphics[width=\textwidth]{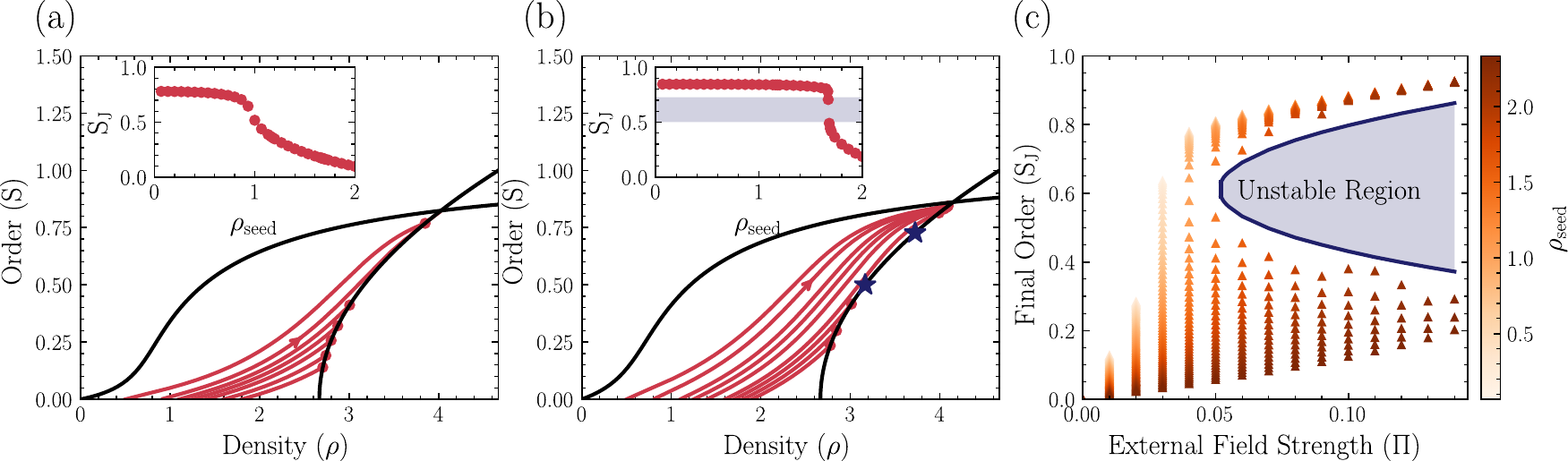}
    \caption{Phase space flows (a,b) bifurcation diagram (c) for $1/\gamma_0=2$ and $r_J=1$. The red lines in (a) and (b) are the trajectories in $(\rho,S)$ space, starting in a disordered state ($S(t=0)=0$) for various values of the initial seeding density $\rho(t=0)=\rho_\mathrm{seed}$. In all cases the trajectories  end on the line of fixed points as the system evolves from the initial disordered state to a state with $\rho=\rho_\mathrm{J}$ and finite alignment ($S\not=0$). All densities on the x-axis are scaled by $\rho_\mathrm{seed}$.
    In (a) we set $\Pi=0.04$ and all points on the lines of fixed points are stable, with  no exceptional points. The final order at jamming depends inversely on seeding density and the order-disorder transition is continuous, as shown in the inset. In (b) for $\Pi=0.07$ there are two  exceptional points (blue stars) bounding the unstable portion of the line of fixed points. For all values of $\rho_\mathrm{seed}$ where the system evolves to an ordered state,  the final order at jamming is independent of seeding density and the order-disorder transition is discontinuous, as shown in the inset. The shaded region in the inset is unstable and is bounded by exceptional points.
    (c)~Bifurcation diagram: each triangle represents a jammed state and the color scale indicates the seeding density. The unstable region is bounded by lines of exceptional points shown in dark blue.}
    \label{fig:phase_space_flows}
\end{figure*}

\begin{figure}
    \centering
    \includegraphics[width=0.8\linewidth]{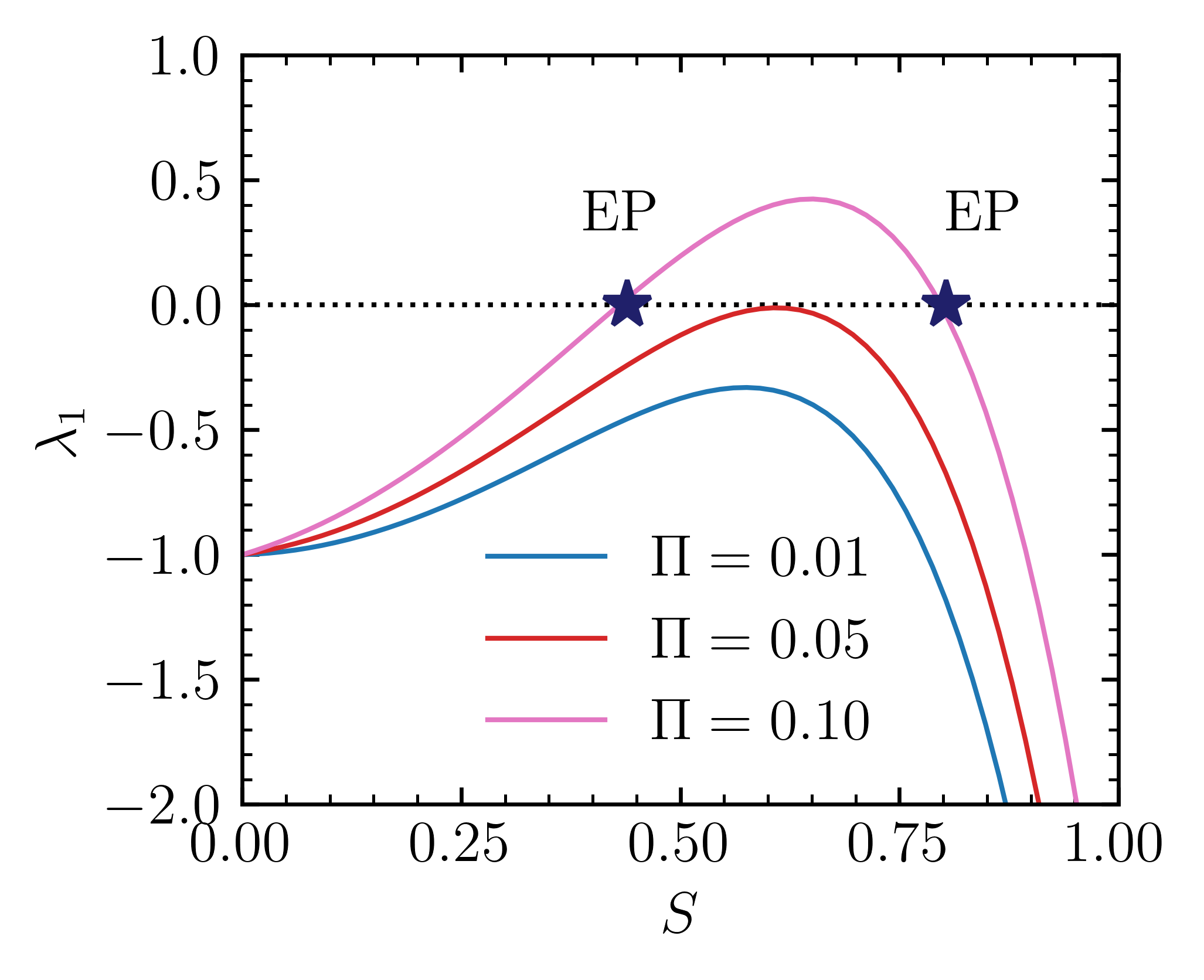}
    \caption{Plot of $\lambda_1(S)$ for $1/\gamma_0=2$ for a few illustrative values of external field strength ($\Pi$). As $\Pi$ crosses a threshold, exceptional points (blue stars) appear.}
    \label{fig:lambd1_plot}
\end{figure}

In the marginal case $r_J=1$, the ODE model exhibits a line of fixed points. This appendix presents a stability analysis that reveals exceptional points where stability switches along this line of fixed points.

For sufficiently large values of $\Pi/\gamma_0$, we can assume that the director angle $\theta$ quickly aligns with the $x$ direction and just consider the coupled dynamics of $S$ and $\rho$.  To examine the stability of the line of fixed points $\rho=\rho_\mathrm{J}(S)$, we consider the Jacobian of this reduced $S-\rho$ system evaluated at the fixed points, given by
\begin{align}
\label{eq:J}
J= \frac{1}{\gamma_0}\begin{bmatrix}
    -A S h(\rho_\mathrm{J}(S), S) & h(\rho_\mathrm{J}(S), S) \\
    \gamma_0 A S & -\gamma_0 
    \end{bmatrix}\,,
\end{align}
with $A=2(\rho_\mathrm{NJ}-\rho_\mathrm{IJ})>0$. 
The eigenvalues and eigenvectors of the Jacobian are given by
\begin{align}
    \lambda_0 &= 0\,, & \mathbf{e}_0 &= (1, AS)\,,\\   
    \lambda_1 &= -1-\frac{A}{\gamma_0}S h(\rho_\mathrm{J}, S) \,, & \mathbf{e}_1 &=\left(-\frac{ h(\rho_\mathrm{J}, S)}{\gamma_0}, 1\right)\,.
\end{align} 

Here $\rho_\mathrm{J}$ depends on $S$ via Eq.~\eqref{eq:rhoJ}, hence eigenvectors and eigenvalues are only functions of $S$. The eigenvector $\mathbf{e}_0$ always points along the tangent to the line of fixed points, implying that perturbations in that direction are marginally stable. The stability of the fixed points in the other direction is controlled by the sign of $\lambda_1(S)$ (see Fig.~\ref{fig:lambd1_plot}). For $\lambda_1<0$, trajectories are attracted towards the line of fixed points, which becomes an absorbing boundary as shown in Fig.~\ref{fig:phase_space_sketch} (green portions). For $\lambda_1>0$, trajectories are repelled from the line of fixed points, corresponding to the orange portion in Fig.~\ref{fig:phase_space_sketch}. When $\lambda_1(S)=0$, both eigenvalues vanish and $\mathbf{e}_1=\left(1/AS,1\right)$, hence the two eigenvectors become colinear and both are tangent to the line of fixed points. The values of $S$ corresponding to $\lambda_1(S)=0$ are denoted by  the blue stars in Fig.~\ref{fig:phase_space_sketch} and correspond to exceptional points. 

The existence of these exceptional points bounding the region of stability of the fixed points $\rho=\rho_\mathrm{J}(S)$ implies a change in the behavior of the system as in this region all trajectories are repelled from the line of fixed points and reach a jammed state with same density and value of $S$,  regardless of the initial
seeding density. 

We can show that the exceptional points, when present, always come in pairs and lie below the point of intersection of the $S$ nullcline and the line of fixed points $\rho=\rho_\mathrm{J}(S)$. This intersection point, denoted by $S^*$, is the solution to $h(\rho_\mathrm{J}(S^*), S^*)=0$. Exceptional points are the solutions of  $\lambda_1(S)=0$, which requires both the following inequalities to be satisfied
\begin{align}
    h(\rho_\mathrm{J}, S) < 0\,, \\
    \label{eq:gamma_crit}
    \frac{1}{\gamma_0} > \frac{-1}{A \, \min[S h(\rho_\mathrm{J}, S)]}\,.
\end{align}
The first condition is satisfied  in the region below the $S$-nullcline for any value of $\Pi$. Thus exceptional points will always be below the intersection point $S^*$. To see that exceptional points always come in pairs, we note that $\lambda_1 (S=0)=-1$ and $\lambda_1(S^*)=-1$. For $\lambda_1(S)$ to be zero between these extrema, the function must cross the $x$-axis at an even number of points, hence the exceptional points always come in pairs. 

\begin{figure}
    \centering
    \includegraphics[width=\linewidth]{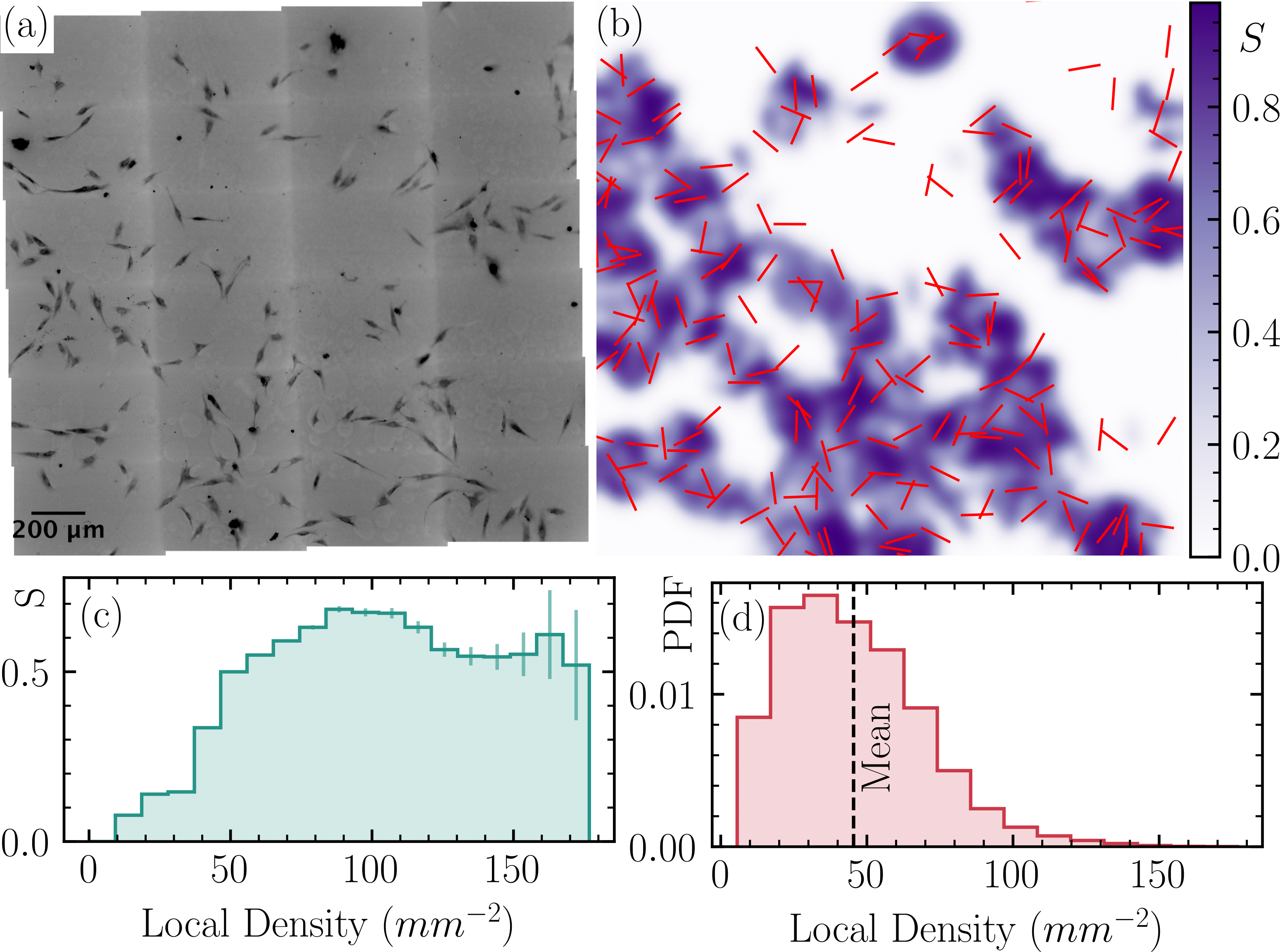}
    \caption{Clustering of cells at low seeding densities. (a) Snapshot of cells on a nematic substrate seeded around $\sim 45 \unit{mm^{-2}}$. (b) Local order (heat-map) of cells calculated using the adaptive-kNN method. Cells are superimposed as red sticks with uniform approximate lengths. (c) Local order as a function of local density shows a correlation between density and order. Error-bars are the standard error of the mean. (d) The histogram of local density shows a long-tailed distribution due to clustering.
    }
    \label{fig:clustering}
\end{figure}

The second condition, Eq.~\eqref{eq:gamma_crit}, where $\gamma_0=\tau_Q/\tau_g$, requires that growth controlled by $\tau_g$ be slower than the remodeling of nematic texture on time scales $\tau_Q$. This sets the threshold for the transition from a disordered to an ordered state. Physically this condition says that if the dynamics of alignment is fast enough compared to growth, the system can order through alignment before jamming.  Alignment is driven by two mechanisms: steric effects captured by the nematic free energy and the external field $\Pi$ that describes the role of substrate anisotropy. Note that increasing  $\Pi$ lowers the  minimum value of $1/{\gamma_0}=\tau_g/\tau_Q$ required for ordering. This is shown in Fig.~\ref{fig:phase_space_flows}, where frames (a) and (b) corresponds to two values of $\Pi$: in (a) for $\Pi=0.04$ there are no exceptional points, while they are present in (b) for $\Pi=0.07$.

Finally, we note that the qualitative form of the phase-space flows does not depend on the exact dependence of $\rho_\mathrm{J}$ on $S$. The results remain the same provided the jamming density $\rho_\mathrm{J}$ is an increasing function of $S$.

\section{Cell clusters}
\label{App:clustering}

In experiments the initial cell density is strongly inhomogeneous, as shown in Fig.~\ref{fig:clustering}. Cells are organized in dense clusters and are aligned with each other within the cluster due to steric effects. The presence of clustering is evident in histograms of local density that exhibit long tails. Local order follows local density. Local order is calculated using an adaptive kNN~\cite{Zhao2022AnalysisEstimation} with cut-off radius, $a\sim \SI{200}{~\mu m}$ and cut-off number of neighbors, $n_c=2$. Essentially near a point if the number of neighbors within the cut-off radius is less than the cut-off number of neighbors, the order is set to zero. The number of neighbors to average, $k$ is chosen adaptively as $k=\lfloor n^{2/3} \rfloor$. This method results in an order field that is consistent with individual cell orientations as seen in Fig.~\ref{fig:clustering}(b). 

To replicate the clustering of cells observed in experiments, we initialize our system  with circular patches of high density (${\sim}\,\SI{180}{mm^{-2}}$) embedded in a low-density background where the density at each grid point is drawn from a uniform distribution between (0--\SI{15}{mm^{-2}}). The above numerical values are chosen from experimental data.
Both the number of clusters and the mean value of the fluctuating background density are assumed to scale linearly with the seeding density. The resulting spatially inhomogeneous density field is normalized so that its mean is the seeding density. Initially, each cluster is uniformly oriented in a randomly chosen direction with scalar order proportional to the local density. The background region is disordered, with random orientation and infinitesimal scalar order.

\begin{figure*}
    \centering
    \includegraphics[width=0.8\linewidth]{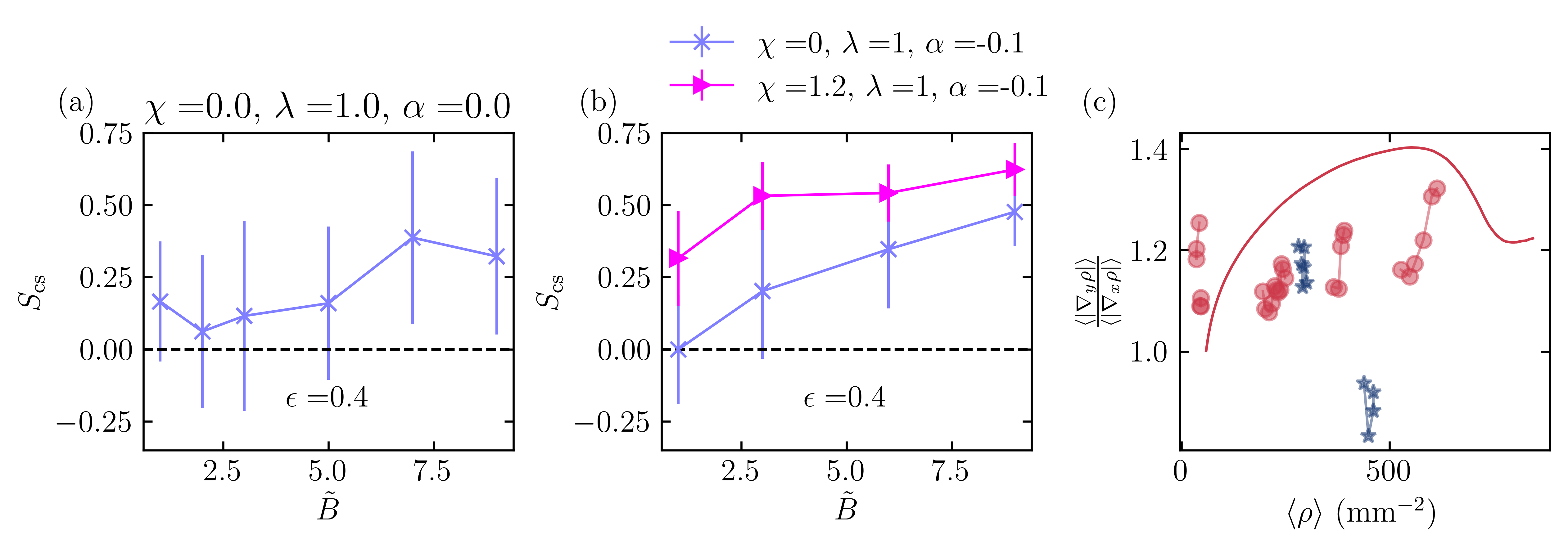}
    \caption{(a-b) Dependence of steady-state cell-substrate order on the nondimensionalized bulk modulus $\tilde{B}$ in simulations with asymmetric friction for friction asymmetry, $\epsilon=0.4$. Lines are drawn to guide the eye.
    (a)~Simulations performed with only flow alignment and no density alignment or activity ($\lambda=1.0, ~ \chi=0,~ \alpha=-0.1$.)
    (b)~Simulations performed with activity and flow alignment but no density alignment ($\lambda=1.0, ~ \chi=0,~ \alpha=-0.1$) show that ``active anchoring'' can lead to substantial order but less compared to explicit anchoring ($\lambda=1.0, ~ \chi=1.2,~ \alpha=-0.1$.) Simulations performed with $\rho_\mathrm{seed}=50~\unit{mm^{-2}}$. Error bars are standard deviations for $n=10$ simulations per data-point.
    (c)~Asymmetry of density gradients in experiments on nematic (red dots) and isotropic (blue stars) substrates as compared to simulations with asymmetric friction (red line).}
    \label{fig:gradient_asymmetry}
\end{figure*}

\section{Signatures of Collective Alignment Mechanism}\label{App:collective_align}

Even though asymmetric friction generates asymmetric flow profiles, flow-alignment alone doesn't seem to drive the ordering to the extent that is seen in experiments. We confirm this in simulations with asymmetric friction where we set activity and density-gradient alignment terms to zero and look at flow-alignment only. We find in Fig.~\ref{fig:gradient_asymmetry}(a) that flow-alignment can drive ordering but it isn't strong. The error-bars are large due to initial conditions being strongly heterogeneous. 

We next look at activity and flow-alignment induced ``active anchoring'' of cells to density gradients that can also, in principle, drive alignment without any explicit anchoring of cells to density gradients. We find in Fig.~\ref{fig:gradient_asymmetry}(b) that indeed activity and flow alignment can drive ordering but it is not as strong as compared to the case with explicit alignment. 

\rvv{It is clear from Fig.~\ref{fig:gradient_asymmetry}(a) that flow-alignment and active pressure can generate cell-substrate alignment.  Deviatoric activity (compare blue points in Fig.~\ref{fig:gradient_asymmetry}(a, b)), can also strengthen cell-substrate order, but its effect is much weaker. This is because the initial state of the system has no nematic order and gradients of active pressure (hence density) dominate over deviatoric active stresses. In the time the system takes to build nematic order, active-pressure-driven alignment is the dominant mechanism. Flows driven by deviatoric active stress cannot be sustained for a sufficient amount of time because the system jams.} 

\rvv{Initial density gradients are crucial for engendering flows, and thus alignment, driven by active pressure. This can be made apparent by considering an isotropic Gaussian patch of high cell density. Assuming $\rho_\mathrm{J}^{}=\rho_\mathrm{J}^{}(0)$, the flow driven by the corresponding initial density gradients is then given by  Eq.~\eqref{eq:Stokes}, which takes the form
\begin{align}
    \label{appendixeq:vel}
    \begin{split}
        v_i &= \frac{(\zeta^{-1})_{ik}}{\rho} \bigg[ -\tilde{B} \nabla_k \rho 
        \\& - \alpha \Theta(\rho_\mathrm{J}^{}(0)-\rho) \left( Q_{kj} \nabla_j\rho + \nabla_j Q_{kj} \right) \bigg],
    \end{split}
\end{align}
where  $\tilde{B} = B e^{\rho/\rho_\mathrm{J}^{}(0)}/\rho_\mathrm{J}^{}(0)$ derives from the gradient of the pressure $p = B e^{\rho/\rho_\mathrm{J}^{}(S)}$ evaluated in the isotropic state $S = 0$. 
The first term on the right hand side of Eq.~\eqref{appendixeq:vel} proportional to $B$ is the contribution from the active pressure due to proliferation. It is finite only in the presence of nonzero density gradients.
The terms proportional to $\alpha$ contribute very little because there is little nematic order in the patch initially; $S\sim 0,\,Q_{ij}\sim 0$. The first two terms give anisotropic diffusion when the velocity is substituted in the evolution equation for the density, but again the main contribution to the initial spreading of density comes from the active pressure. }

\rvv{Active pressure gradients can generate cell-substrate alignment through both the $\chi$-term and the flow-alignment term (proportional to $\lambda$). Here we compare their relative importance. To do so we consider the effect of the first term in Eq.~\eqref{appendixeq:vel} on the spreading of a Gaussian density mound (of initial radius $r_0$ and max density $\rho_0$). The root-mean-square (RMS) spread in a time $\tau_g$ can be calculated by ignoring birth/death and considering only diffusion. The RMS spreads are, $(\Delta x)^2 \sim \tilde B\tau_g/\zeta_{xx}$ and $ (\Delta y)^2 \sim \tilde B\tau_g/\zeta_{yy}$. Since $\zeta_{yy}>\zeta_{xx}$, the RMS spread in $x$ is higher than the spread in $y$. Thus in a time $\sim\tau_g$ the density gradients become anisotropic and of order $|\nabla_x \rho| \sim \rho_0/(r_0+\Delta x)$ and $|\nabla_y \rho| \sim \rho_0/(r_0+\Delta y) $. Since the spread is higher in $x$ than in $y$, we have $|\nabla_x\rho|<|\nabla_y\rho|$. With these estimates for density gradients we can determine the contribution of gradient alignment to cell-substrate order
\begin{equation}
    [\Delta S_\mathrm{cs}]_\chi \sim \frac{\chi \tau_g}{2} [(\nabla_x\rho)^2-(\nabla_y\rho)^2] \sim \epsilon\,\frac{\chi\tau_g\rho_0^2}{r_0^3}\sqrt{\frac{\tilde B\tau_g}{\zeta_0}}\;.
\end{equation}
For flow alignment, we have
\begin{equation}
    [\Delta S_\mathrm{cs}]_\lambda \sim \lambda(\rho) [\nabla_xv_x-\nabla_yv_y]\tau_g\;.
\end{equation}
The velocities are approximated as
\begin{equation}
    v_x \sim \dfrac{B}{\zeta_{xx}\rho_\mathrm{J}^{}(0)}\dfrac{\nabla_x\rho}{\rho}\,, \quad v_y \sim \dfrac{B}{\zeta_{yy}\rho_\mathrm{J}^{}(0)}\dfrac{\nabla_y\rho}{\rho}\;,
\end{equation}
which gives
\begin{align}
    \begin{split}
        [\Delta S_\mathrm{cs}]_\lambda &\sim \lambda(\rho) \left(\frac{\tilde B\tau_g}{r_0^2}\right)^{\!3/2} \left[ \frac{1}{\zeta_{xx}^{3/2}} -\frac{1}{\zeta_{yy}^{3/2}}\right]
        \\ & \sim 3\epsilon\,\lambda(\rho) \left(\frac{\tilde B\tau_g}{\zeta_0 r_0^2}\right)^{\!3/2}\;.
    \end{split}
\end{align}
Hence both flow alignment ($\sim\lambda$) and density-gradient driven alignment ($\sim \chi$) contribute to ordering to first order in the friction anisotropy, $\epsilon$, but the $\chi$-term scales as $\rho_0^2$ while the flow-alignment contribution scales as $\rho_\mathrm{J}^{}-\rho$ due to the dependence of $\lambda$ on $\rho$. For the initial Gaussian mound where densities are generally above the I--N transition density but less than the jamming density, the $\chi$-term will be stronger than the flow-alignment term. We once again stress that here, unlike in previous literature, flow-alignment is generated by anisotropic density gradients not by  deviatoric active stresses.}

Next we looked for asymmetry in density gradients along the $x$- versus $y$-direction in experiments as well as simulations with asymmetric friction. We find that in experiments on nematic substrates the asymmetry in gradients is consistently present at all densities (see red dots in Fig.~\ref{fig:gradient_asymmetry}(c)). However, in experiments on isotropic substrates the asymmetry averages out to zero across experiments (see one set of blue stars above 1 and another below 1 in Fig.~\ref{fig:gradient_asymmetry}(c)). In simulations with asymmetric friction ($\epsilon=0.4$) the asymmetry gradually increases but is consistently higher than experiments.

\begin{figure}
    \centering
    \includegraphics[width=\linewidth]{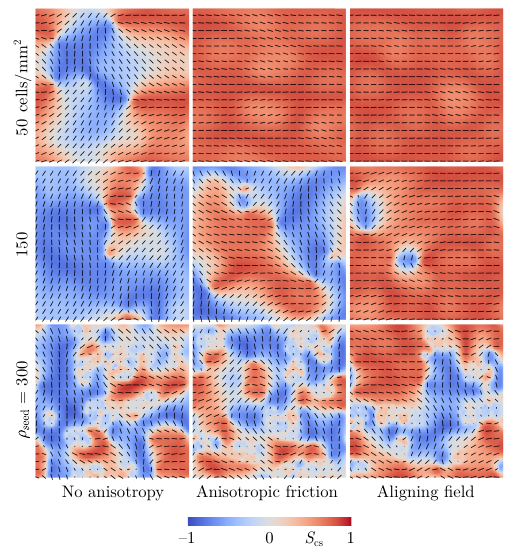}
    \caption{\rv{Snapshots at jamming of the cell substrate order field (colormaps) and nematic orientations (black directors, length scaled by nematic order magnitude). The area fraction of substrate-aligned regions (red) reduces with increasing seeding density (see Fig.~\ref{fig:Scs-rho_seed} for a quantification). For the highest seeding density $\rho_\mathrm{seed} = \SI{300}{cells/mm^2}$, local nematic order is also low in large regions as indicated by short directors.}}
    \label{fig:final_snapshots}
\end{figure}

\section{External alignment field}
\label{app:external-field}

\rv{The main text presents simulations of the spatially extended model with anisotropic friction, which we show leads to collective cell alignment along the low-friction direction. 
In this appendix, we compare this model variant to one where the orientational bias imparted on the cells by the substrate is implemented via an ``external field,'' analogously to the ODE model.}

\begin{figure}
    \centering
    \includegraphics[width=\linewidth]{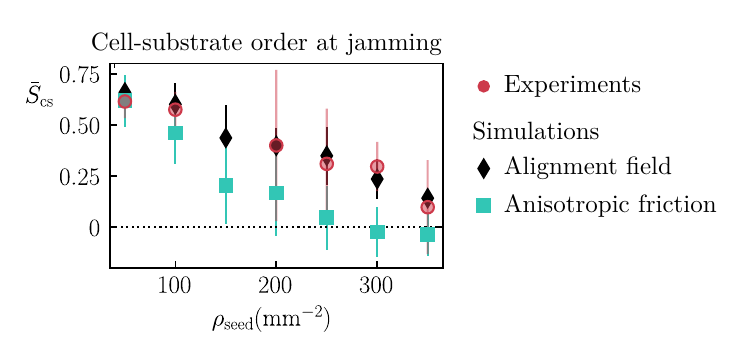}
    \caption{\rv{The cell-substrate order at jamming obtained from simulations with an external aligning field (black dots) and asymmetric friction (cyan squares) decreases with increasing seeding density. Red dots are experiments on a nematic substrate.}}
    \label{fig:Scs-rho_seed}
\end{figure}

\rv{We find that both model variants reproduce the growth of local and global order with mean density and the dependence of cell-substrate order on the initial seeding density, as shown in Fig.~\ref{fig:sim_results_both}(a-c), Videos~($1$--$3$), and Fig.~\ref{fig:final_snapshots}. In particular, both variants capture the experimental observation that at intermediate density (${\sim}\SI{200}{cells/mm^{2}}$) there is appreciable local order [see Fig.~\ref{fig:sim_results_both}(b)] but little cell-substrate order [see Fig.~\ref{fig:sim_results_both}(a)]. Moreover, the spatially extended model also reproduces the result obtained with the ODE that the cell-substrate order at jamming decreases with increasing seeding density (Fig.~\ref{fig:Scs-rho_seed}). The model with sn external field does not, however, explain the mechanisms that drive collective alignment to the substrate. }

\clearpage

\bibliography{references}

@article{DellArciprete2018ANematic,
    title = {{A growing bacterial colony in two dimensions as an active nematic}},
    year = {2018},
    journal = {Nature Communications 2018 9:1},
    author = {Dell’Arciprete, D. and Blow, M. L. and Brown, A. T. and Farrell, F. D.C. and Lintuvuori, J. S. and McVey, A. F. and Marenduzzo, D. and Poon, W. C.K.},
    number = {1},
    month = {10},
    pages = {1--9},
    volume = {9},
    publisher = {Nature Publishing Group},
    url = {https://www.nature.com/articles/s41467-018-06370-3},
    doi = {10.1038/s41467-018-06370-3},
    issn = {2041-1723},
    pmid = {30305618}
}

@article{Doostmohammadi2018ActiveNematics,
    title = {{Active nematics}},
    year = {2018},
    journal = {Nature Communications 2018 9:1},
    author = {Doostmohammadi, Amin and Ign{\'{e}}s-Mullol, Jordi and Yeomans, Julia M. and Sagu{\'{e}}s, Francesc},
    number = {1},
    month = {8},
    pages = {1--13},
    volume = {9},
    publisher = {Nature Publishing Group},
    url = {https://www.nature.com/articles/s41467-018-05666-8},
    doi = {10.1038/s41467-018-05666-8},
    issn = {2041-1723},
    pmid = {30131558}
}

@article{Balasubramaniam2022ActiveMorphogenesis,
    title = {{Active nematics across scales from cytoskeleton organization to tissue morphogenesis}},
    year = {2022},
    journal = {Current Opinion in Genetics {\&} Development},
    author = {Balasubramaniam, Lakshmi and M{\`{e}}ge, René Marc and Ladoux, Benoît},
    month = {4},
    pages = {101897},
    volume = {73},
    publisher = {Elsevier Current Trends},
    doi = {10.1016/J.GDE.2021.101897},
    issn = {0959-437X},
    pmid = {35063879}
}

@article{Thijssen2020ActiveParameter,
    title = {{Active nematics with anisotropic friction: the decisive role of the flow aligning parameter}},
    year = {2020},
    journal = {Soft Matter},
    author = {Thijssen, Kristian and Metselaar, Luuk and Yeomans, Julia M. and Doostmohammadi, Amin},
    number = {8},
    month = {2},
    pages = {2065--2074},
    volume = {16},
    publisher = {Royal Society of Chemistry},
    url = {https://pubs.rsc.org/en/content/articlehtml/2020/sm/c9sm01963d https://pubs.rsc.org/en/content/articlelanding/2020/sm/c9sm01963d},
    doi = {10.1039/C9SM01963D},
    issn = {17446848},
    arxivId = {1910.00331}
}

@article{Endresen2024Actuation3D,
    title = {{Actuation of Cell Sheets in 3D}},
    year = {2024},
    journal = {arxiv},
    author = {Endresen, Kirsten and Murali, Aniruddh and Serra, Francesca},
    month = {11},
    pages = {2411.17834},
    url = {https://arxiv.org/abs/2411.17834v1},
    arxivId = {2411.17834}
}

@article{Jacques2023AgingRemodeling,
    title = {{Aging and freezing of active nematic dynamics of cancer-associated fibroblasts by fibronectin matrix remodeling}},
    year = {2023},
    journal = {bioRxiv},
    author = {Jacques, Cécile and Ackermann, Joseph and Bell, Samuel and Hallopeau, Clément and Gonzalez, Carlos Perez- and Balasubramaniam, Lakshmi and Trepat, Xavier and Ladoux, Benoît and Maitra, Ananyo and Voituriez, Raphael and Vignjevic, Danijela Matic},
    month = {11},
    pages = {2023.11.22.568216},
    publisher = {Cold Spring Harbor Laboratory},
    url = {https://www.biorxiv.org/content/10.1101/2023.11.22.568216v1 https://www.biorxiv.org/content/10.1101/2023.11.22.568216v1.abstract},
    institution = {bioRxiv},
    doi = {10.1101/2023.11.22.568216}
}

@article{Zhao2022AnalysisEstimation,
    title = {{Analysis of KNN Density Estimation}},
    year = {2022},
    journal = {IEEE Transactions on Information Theory},
    author = {Zhao, Puning and Lai, Lifeng},
    number = {12},
    month = {12},
    pages = {7971--7995},
    volume = {68},
    publisher = {Institute of Electrical and Electronics Engineers Inc.},
    doi = {10.1109/TIT.2022.3195870},
    issn = {15579654},
    arxivId = {2010.00438}
}

@article{Baron2023AnisotropicFields,
    title = {{Anisotropic particle multiphase equilibria in nonuniform fields}},
    year = {2023},
    journal = {Journal of Chemical Physics},
    author = {Baron, Philippe B. and Hendley, Rachel S. and Bevan, Michael A.},
    number = {12},
    month = {9},
    pages = {124902},
    volume = {159},
    publisher = {American Institute of Physics Inc.},
    url = {/aip/jcp/article/159/12/124902/2912116/Anisotropic-particle-multiphase-equilibria-in},
    doi = {10.1063/5.0169659/18134720/124902{\_}1{\_}5.0169659.AM.PDF},
    issn = {10897690},
    pmid = {38127375}
}

@article{Volfson2008BiomechanicalPopulations,
    title = {{Biomechanical ordering of dense cell populations}},
    year = {2008},
    journal = {Proceedings of the National Academy of Sciences of the United States of America},
    author = {Volfson, Dmitri and Cookson, Scott and Hasty, Jeff and Tsimring, Lev S.},
    number = {40},
    month = {10},
    pages = {15346--15351},
    volume = {105},
    publisher = {{\\\}National Academy of Sciences},
    url = {https://www.pnas.org/doi/abs/10.1073/pnas.0706805105},
    doi = {10.1073/PNAS.0706805105/SUPPL{\_}FILE/0706805105SI.PDF},
    issn = {00278424},
    pmid = {18832176},
    keywords = {Bacteria, Biofilms, Microfluidics, Nematodynamics}
}

@article{Blow2014BiphasicNematicsb,
    title = {{Biphasic, lyotropic, active nematics}},
    year = {2014},
    journal = {Physical Review Letters},
    author = {Blow, Matthew L. and Thampi, Sumesh P. and Yeomans, Julia M.},
    number = {24},
    month = {12},
    pages = {248303},
    volume = {113},
    publisher = {American Physical Society},
    url = {https://journals.aps.org/prl/abstract/10.1103/PhysRevLett.113.248303},
    doi = {10.1103/PHYSREVLETT.113.248303/SI.PDF},
    issn = {10797114},
    pmid = {25541809},
    arxivId = {1407.7493}
}

@article{Erdogan2017Cancer-associatedFibronectin,
    title = {{Cancer-associated fibroblasts promote directional cancer cell migration by aligning fibronectin}},
    year = {2017},
    journal = {The Journal of cell biology},
    author = {Erdogan, Begum and Ao, Mingfang and White, Lauren M. and Means, Anna L. and Brewer, Bryson M. and Yang, Lijie and Washington, M. Kay and Shi, Chanjuan and Franco, Omar E. and Weaver, Alissa M. and Hayward, Simon W. and Li, Deyu and Webb, Donna J.},
    number = {11},
    month = {11},
    pages = {3799--3816},
    volume = {216},
    publisher = {J Cell Biol},
    url = {https://pubmed.ncbi.nlm.nih.gov/29021221/},
    doi = {10.1083/JCB.201704053},
    issn = {1540-8140},
    pmid = {29021221}
}

@article{Huang2025Cell-SheetForces,
    title = {{Cell-Sheet Shape Transformation by Internally-Driven, Oriented Forces}},
    year = {2025},
    journal = {Advanced Materials},
    author = {Huang, Junrou and Chen, Juan and Luo, Yimin},
    number = {20},
    month = {5},
    pages = {2416624},
    volume = {37},
    publisher = {John Wiley {\&} Sons, Ltd},
    url = {https://onlinelibrary.wiley.com/doi/full/10.1002/adma.202416624 https://onlinelibrary.wiley.com/doi/abs/10.1002/adma.202416624 https://advanced.onlinelibrary.wiley.com/doi/10.1002/adma.202416624},
    doi = {10.1002/ADMA.202416624},
    issn = {1521-4095},
    pmid = {40165759}
}

@article{Stringer2025Cellpose3:Segmentation,
    title = {{Cellpose3: one-click image restoration for improved cellular segmentation}},
    year = {2025},
    journal = {Nature Methods 2025 22:3},
    author = {Stringer, Carsen and Pachitariu, Marius},
    number = {3},
    month = {2},
    pages = {592--599},
    volume = {22},
    publisher = {Nature Publishing Group},
    url = {https://www.nature.com/articles/s41592-025-02595-5},
    doi = {10.1038/s41592-025-02595-5},
    issn = {1548-7105}
}

@article{Stringer2020Cellpose:Segmentation,
    title = {{Cellpose: a generalist algorithm for cellular segmentation}},
    year = {2020},
    journal = {Nature Methods 2020 18:1},
    author = {Stringer, Carsen and Wang, Tim and Michaelos, Michalis and Pachitariu, Marius},
    number = {1},
    month = {12},
    pages = {100--106},
    volume = {18},
    publisher = {Nature Publishing Group},
    url = {https://www.nature.com/articles/s41592-020-01018-x},
    doi = {10.1038/s41592-020-01018-x},
    issn = {1548-7105},
    pmid = {33318659}
}

@article{Puliafito2012CollectiveInhibition,
    title = {{Collective and single cell behavior in epithelial contact inhibition}},
    year = {2012},
    journal = {Proceedings of the National Academy of Sciences of the United States of America},
    author = {Puliafito, Alberto and Hufnagel, Lars and Neveu, Pierre and Streichan, Sebastian and Sigal, Alex and Fygenson, D. Kuchnir and Shraiman, Boris I.},
    number = {3},
    month = {1},
    pages = {739--744},
    volume = {109},
    url = {www.pnas.org/lookup/suppl/},
    doi = {10.1073/PNAS.1007809109/-/DCSUPPLEMENTAL/PNAS.1007809109{\_}SI.PDF},
    issn = {00278424},
    pmid = {22228306},
    arxivId = {1112.0465}
}

@article{Wu2025CollectivePerspective,
    title = {{Collective cell migration across scales: A systems perspective}},
    year = {2025},
    journal = {Seminars in Cell {\&} Developmental Biology},
    author = {Wu, Zimeng and Wong, Mie},
    month = {9},
    pages = {103628},
    volume = {173},
    publisher = {Academic Press},
    doi = {10.1016/J.SEMCDB.2025.103628},
    issn = {1084-9521}
}

@article{Ray2024ConfinementEmbryogenesis,
    title = {{Confinement promotes nematic alignment of spindle-shaped cells during Drosophila embryogenesis}},
    year = {2024},
    journal = {Development (Cambridge)},
    author = {Ray, Tirthankar and Shi, Damo and Harris, Tony J.C.},
    number = {13},
    month = {7},
    volume = {151},
    publisher = {Company of Biologists Ltd},
    url = {https://dx.doi.org/10.1242/dev.202577},
    doi = {10.1242/DEV.202577/VIDEO-4},
    issn = {14779129},
    pmid = {38864272}
}

@article{You2021Confinement-inducedColonies,
    title = {{Confinement-induced self-organization in growing bacterial colonies}},
    year = {2021},
    journal = {Science Advances},
    author = {You, Zhihong and Pearce, Daniel J.G. and Giomi, Luca},
    number = {4},
    month = {1},
    volume = {7},
    publisher = {American Association for the Advancement of Science},
    doi = {10.1126/SCIADV.ABC8685},
    issn = {23752548},
    pmid = {33523940},
    arxivId = {2004.14890}
}

@article{Sarkar2023CrisscrossSheets,
    title = {{Crisscross multilayering of cell sheets}},
    year = {2023},
    journal = {PNAS Nexus},
    author = {Sarkar, Trinish and Yashunsky, Victor and Br{\'{e}}zin, Louis and Mercader, Carles Blanch and Aryaksama, Thibault and Lacroix, Mathilde and Risler, Thomas and Joanny, Jean François and Silberzan, Pascal},
    number = {3},
    month = {3},
    pages = {1--11},
    volume = {2},
    publisher = {Oxford Academic},
    url = {https://dx.doi.org/10.1093/pnasnexus/pgad034},
    doi = {10.1093/PNASNEXUS/PGAD034},
    issn = {27526542}
}

@article{Lacroix2024EmergenceGuidance,
    title = {{Emergence of bidirectional cell laning from collective contact guidance}},
    year = {2024},
    journal = {Nature Physics 2024 20:8},
    author = {Lacroix, Mathilde and Smeets, Bart and Blanch-Mercader, Carles and Bell, Samuel and Giuglaris, Caroline and Chen, Hsiang Ying and Prost, Jacques and Silberzan, Pascal},
    number = {8},
    month = {5},
    pages = {1324--1331},
    volume = {20},
    publisher = {Nature Publishing Group},
    url = {https://www.nature.com/articles/s41567-024-02510-3},
    doi = {10.1038/s41567-024-02510-3},
    issn = {1745-2481}
}

@article{Adar2024Environment-StoredRemodeling,
    title = {{Environment-Stored Memory in Active Nematics and Extra-Cellular Matrix Remodeling}},
    year = {2024},
    journal = {Physical Review Letters},
    author = {Adar, Ram M. and Joanny, Jean-François},
    number = {11},
    month = {9},
    pages = {118402},
    volume = {133},
    doi = {10.1103/PhysRevLett.133.118402},
    issn = {0031-9007}
}

@article{Schindelin2012Fiji:Analysis,
    title = {{Fiji: an open-source platform for biological-image analysis}},
    year = {2012},
    journal = {Nature Methods 2012 9:7},
    author = {Schindelin, Johannes and Arganda-Carreras, Ignacio and Frise, Erwin and Kaynig, Verena and Longair, Mark and Pietzsch, Tobias and Preibisch, Stephan and Rueden, Curtis and Saalfeld, Stephan and Schmid, Benjamin and Tinevez, Jean Yves and White, Daniel James and Hartenstein, Volker and Eliceiri, Kevin and Tomancak, Pavel and Cardona, Albert},
    number = {7},
    month = {6},
    pages = {676--682},
    volume = {9},
    publisher = {Nature Publishing Group},
    url = {https://www.nature.com/articles/nmeth.2019},
    doi = {10.1038/nmeth.2019},
    issn = {1548-7105},
    pmid = {22743772}
}

@article{Resnick2003FluidWorse,
    title = {{Fluid shear stress and the vascular endothelium: for better and for worse}},
    year = {2003},
    journal = {Progress in Biophysics and Molecular Biology},
    author = {Resnick, Nitzan and Yahav, Hava and Shay-Salit, Ayelet and Shushy, Moran and Schubert, Shay and Zilberman, Limor Chen Michal and Wofovitz, Efrat},
    number = {3},
    month = {4},
    pages = {177--199},
    volume = {81},
    publisher = {Pergamon},
    doi = {10.1016/S0079-6107(02)00052-4},
    issn = {0079-6107},
    pmid = {12732261}
}

@article{Ranft2010FluidizationApoptosis,
    title = {{Fluidization of tissues by cell division and apoptosis}},
    year = {2010},
    journal = {Proceedings of the National Academy of Sciences of the United States of America},
    author = {Ranft, Jonas and Basan, Markus and Elgeti, Jens and Joanny, Jean François and Prost, Jacques and J{\"{u}}licher, Frank},
    number = {49},
    month = {12},
    pages = {20863--20868},
    volume = {107},
    publisher = {National Academy of Sciences},
    url = {https://www.pnas.org/doi/abs/10.1073/pnas.1011086107},
    doi = {10.1073/PNAS.1011086107/SUPPL{\_}FILE/SM01.MOV},
    issn = {00278424},
    pmid = {21078958}
}

@article{Heisenberg2013ForcesPatterning,
    title = {{Forces in Tissue Morphogenesis and Patterning}},
    year = {2013},
    journal = {Cell},
    author = {Heisenberg, Carl Philipp and Bella{\"{i}}che, Yohanns},
    number = {5},
    month = {5},
    pages = {948--962},
    volume = {153},
    publisher = {Cell Press},
    doi = {10.1016/J.CELL.2013.05.008},
    issn = {0092-8674},
    pmid = {23706734}
}

@article{Casale2021GeometricalTissues,
    title = {{Geometrical confinement controls cell, ECM and vascular network alignment during the morphogenesis of 3D bioengineered human connective tissues}},
    year = {2021},
    journal = {Acta Biomaterialia},
    author = {Casale, Costantino and Imparato, Giorgia and Mazio, Claudia and Netti, Paolo A. and Urciuolo, Francesco},
    month = {9},
    pages = {341--354},
    volume = {131},
    publisher = {Elsevier},
    doi = {10.1016/J.ACTBIO.2021.06.022},
    issn = {1742-7061},
    pmid = {34144214}
}

@article{Angelini2011Glass-likeMigration,
    title = {{Glass-like dynamics of collective cell migration}},
    year = {2011},
    journal = {Proceedings of the National Academy of Sciences of the United States of America},
    author = {Angelini, Thomas E. and Hannezo, Edouard and Trepatc, Xavier and Marquez, Manuel and Fredberg, Jeffrey J. and Weitz, David A.},
    number = {12},
    month = {3},
    pages = {4714--4719},
    volume = {108},
    doi = {10.1073/PNAS.1010059108/-/DCSUPPLEMENTAL/PNAS.1010059108{\_}SI.PDF},
    issn = {00278424},
    pmid = {21321233}
}

@article{Guillamat2025GuidanceSurfaces,
    title = {{Guidance of cellular nematics into shape-programmable living surfaces}},
    year = {2025},
    journal = {bioRxiv},
    author = {Guillamat, Pau and Mirza, Waleed and Bal, Pradeep K. and G{\'{o}}mez-Gonz{\'{a}}lez, Manuel and Roca-Cusachs, Pere and Arroyo, Marino and Trepat, Xavier},
    month = {6},
    pages = {2025.06.27.660992},
    publisher = {Cold Spring Harbor Laboratory},
    url = {https://www.biorxiv.org/content/10.1101/2025.06.27.660992v1 https://www.biorxiv.org/content/10.1101/2025.06.27.660992v1.abstract},
    institution = {bioRxiv},
    doi = {10.1101/2025.06.27.660992}
}

@misc{Parmar2025Https://github.com/toshi-physics/pyPSS-proliferating-nematic,
    title = {{https://github.com/toshi-physics/pyPSS-proliferating-nematic}},
    year = {2025},
    booktitle = {https://github.com/toshi-physics/pyPSS-proliferating-nematic{\#}},
    author = {Parmar, Toshi},
    url = {https://github.com/toshi-physics/pyPSS-proliferating-nematic#}
}

@article{Zhao2025IntegerMonolayers,
    title = {{Integer topological defects offer a methodology to quantify and classify active cell monolayers}},
    year = {2025},
    journal = {Nature Communications 2025 16:1},
    author = {Zhao, Zihui and Li, He and Yao, Yisong and Zhao, Yongfeng and Serra, Francesca and Kawaguchi, Kyogo and Zhang, Hepeng and Sano, Masaki},
    number = {1},
    month = {3},
    pages = {1--11},
    volume = {16},
    publisher = {Nature Publishing Group},
    url = {https://www.nature.com/articles/s41467-025-57783-w},
    doi = {10.1038/s41467-025-57783-w},
    issn = {2041-1723}
}

@article{Guillamat2022IntegerMorphogenesis,
    title = {{Integer topological defects organize stresses driving tissue morphogenesis}},
    year = {2022},
    journal = {Nature Materials 2022 21:5},
    author = {Guillamat, Pau and Blanch-Mercader, Carles and Pernollet, Guillaume and Kruse, Karsten and Roux, Aurélien},
    number = {5},
    month = {2},
    pages = {588--597},
    volume = {21},
    publisher = {Nature Publishing Group},
    url = {https://www.nature.com/articles/s41563-022-01194-5},
    doi = {10.1038/s41563-022-01194-5},
    issn = {1476-4660},
    pmid = {35145258}
}

@article{Lawson-Keister2021JammingTissues,
    title = {{Jamming and arrest of cell motion in biological tissues}},
    year = {2021},
    journal = {Current Opinion in Cell Biology},
    author = {Lawson-Keister, Elizabeth and Manning, M. Lisa},
    month = {10},
    pages = {146--155},
    volume = {72},
    publisher = {Elsevier Current Trends},
    doi = {10.1016/J.CEB.2021.07.011},
    issn = {0955-0674},
    pmid = {34461581},
    arxivId = {2102.11255}
}

@book{Biau2015LecturesMethod,
    title = {{Lectures on the Nearest Neighbor Method}},
    year = {2015},
    author = {Biau, Gérard and Devroye, Luc},
    series = {Springer Series in the Data Sciences},
    publisher = {Springer International Publishing},
    url = {http://link.springer.com/10.1007/978-3-319-25388-6},
    address = {Cham},
    isbn = {978-3-319-25386-2},
    doi = {10.1007/978-3-319-25388-6}
}

@article{Martella2019LiquidAlignment,
    title = {{Liquid Crystal-Induced Myoblast Alignment}},
    year = {2019},
    journal = {Advanced Healthcare Materials},
    author = {Martella, Daniele and Pattelli, Lorenzo and Matassini, Camilla and Ridi, Francesca and Bonini, Massimo and Paoli, Paolo and Baglioni, Piero and Wiersma, Diederik S. and Parmeggiani, Camilla},
    number = {3},
    month = {2},
    pages = {1801489},
    volume = {8},
    publisher = {John Wiley {\&} Sons, Ltd},
    url = {https://onlinelibrary.wiley.com/doi/full/10.1002/adhm.201801489 https://onlinelibrary.wiley.com/doi/abs/10.1002/adhm.201801489 https://advanced.onlinelibrary.wiley.com/doi/10.1002/adhm.201801489},
    doi = {10.1002/ADHM.201801489},
    issn = {2192-2659},
    pmid = {30605262}
}

@article{Luo2023Molecular-scaleMonolayers,
    title = {{Molecular-scale substrate anisotropy, crowding and division drive collective behaviours in cell monolayers}},
    year = {2023},
    journal = {Journal of the Royal Society Interface},
    author = {Luo, Yimin and Gu, Mengyang and Park, Minwook and Fang, Xinyi and Kwon, Younghoon and Urue{\~{n}}a, Juan Manuel and Read De Alaniz, Javier and Helgeson, Matthew E. and Marchetti, Cristina M. and Valentine, Megan T.},
    number = {204},
    month = {7},
    volume = {20},
    publisher = {{\\\}The Royal Society{\\\}},
    url = {https://royalsocietypublishing.org/doi/10.1098/rsif.2023.0160},
    doi = {10.1098/RSIF.2023.0160},
    issn = {17425662},
    pmid = {37403487},
    keywords = {active nematics, anisotropy, cell migration}
}

@article{Bi2016Motility-drivenTissues,
    title = {{Motility-driven glass and jamming transitions in biological tissues}},
    year = {2016},
    journal = {Physical Review X},
    author = {Bi, Dapeng and Yang, Xingbo and Marchetti, M. Cristina and Manning, M. Lisa},
    number = {2},
    month = {4},
    pages = {021011},
    volume = {6},
    publisher = {American Physical Society},
    url = {https://journals.aps.org/prx/abstract/10.1103/PhysRevX.6.021011},
    doi = {10.1103/PHYSREVX.6.021011/FLUID{\_}TISSUE{\_}V0{\_}0.2{\_}P0{\_}3.8{\_}DR{\_}0.1.MP4},
    issn = {21603308},
    arxivId = {1509.06578}
}

@article{Kemkemer2000NematicCells,
    title = {{Nematic order-disorder state transition in a liquid crystal analogue formed by oriented and migrating amoeboid cells}},
    year = {2000},
    journal = {Eur. Phys. J. E},
    author = {Kemkemer, R and Teichgr{\"{a}}ber, V and Schrank-Kaufmann, S and Kaufmann, D and Gruler, H},
    pages = {101--110},
    volume = {3}
}

@article{Revignas2023OnCrystals,
    title = {{On the elusive saddle-splay and splay-bend elastic constants of nematic liquid crystals}},
    year = {2023},
    journal = {Journal of Chemical Physics},
    author = {Revignas, Davide and Ferrarini, Alberta},
    number = {3},
    month = {7},
    pages = {34905},
    volume = {159},
    publisher = {American Institute of Physics Inc.},
    url = {/aip/jcp/article/159/3/034905/2903245/On-the-elusive-saddle-splay-and-splay-bend-elastic},
    doi = {10.1063/5.0153831/2903245},
    issn = {10897690},
    pmid = {37470424}
}

@article{Li2017OnFibroblasts,
    title = {{On the mechanism of long-range orientational order of fibroblasts}},
    year = {2017},
    journal = {Proceedings of the National Academy of Sciences of the United States of America},
    author = {Li, Xuefei and Balagam, Rajesh and He, Ting Fang and Lee, Peter P. and Igoshin, Oleg A. and Levine, Herbert},
    number = {34},
    month = {8},
    pages = {8974--8979},
    volume = {114},
    publisher = {National Academy of Sciences},
    url = {https://www.pnas.org/doi/abs/10.1073/pnas.1707210114},
    doi = {10.1073/PNAS.1707210114/SUPPL{\_}FILE/PNAS.1707210114.SAPP.PDF},
    issn = {10916490},
    pmid = {28784754}
}

@article{Duclos2014PerfectCells,
    title = {{Perfect nematic order in confined monolayers of spindle-shaped cells}},
    year = {2014},
    journal = {Soft Matter},
    author = {Duclos, G. and Garcia, S. and Yevick, H. G. and Silberzan, P.},
    number = {14},
    month = {3},
    pages = {2346--2353},
    volume = {10},
    publisher = {The Royal Society of Chemistry},
    url = {https://pubs.rsc.org/en/content/articlehtml/2014/sm/c3sm52323c https://pubs.rsc.org/en/content/articlelanding/2014/sm/c3sm52323c},
    doi = {10.1039/C3SM52323C},
    issn = {1744-6848},
    pmid = {24623001}
}

@article{Bautista-Carbajal2014PhaseEllipses,
    title = {{Phase diagram of two-dimensional hard ellipses}},
    year = {2014},
    journal = {Journal of Chemical Physics},
    author = {Bautista-Carbajal, Gustavo and Odriozola, Gerardo},
    number = {20},
    month = {5},
    pages = {204502},
    volume = {140},
    publisher = {American Institute of Physics Inc.},
    url = {/aip/jcp/article/140/20/204502/74145/Phase-diagram-of-two-dimensional-hard-ellipses},
    doi = {10.1063/1.4878411/74145},
    issn = {00219606},
    arxivId = {1405.2236}
}

@article{Garcia2015PhysicsMonolayer,
    title = {{Physics of active jamming during collective cellular motion in a monolayer}},
    year = {2015},
    journal = {Proceedings of the National Academy of Sciences of the United States of America},
    author = {Garcia, Simon and Hannezo, Edouard and Elgeti, Jens and Joanny, Jean François and Silberzan, Pascal and Gov, Nir S.},
    number = {50},
    month = {12},
    pages = {15314--15319},
    volume = {112},
    publisher = {National Academy of Sciences},
    doi = {10.1073/PNAS.1510973112/VIDEO-3},
    issn = {10916490},
    pmid = {26627719}
}

@article{Brezin2022SpontaneousDefects,
    title = {{Spontaneous flow created by active topological defects}},
    year = {2022},
    journal = {The European Physical Journal E 2022 45:4},
    author = {Br{\'{e}}zin, Louis and Risler, Thomas and Joanny, Jean Francois},
    number = {4},
    month = {4},
    pages = {1--16},
    volume = {45},
    publisher = {Springer},
    url = {https://link.springer.com/article/10.1140/epje/s10189-022-00186-2},
    doi = {10.1140/EPJE/S10189-022-00186-2},
    issn = {1292-895X},
    pmid = {35389081},
    arxivId = {2202.00646}
}

@article{Sanchez2012SpontaneousMatter,
    title = {{Spontaneous motion in hierarchically assembled active matter}},
    year = {2012},
    journal = {Nature 2012 491:7424},
    author = {Sanchez, Tim and Chen, Daniel T.N. and Decamp, Stephen J. and Heymann, Michael and Dogic, Zvonimir},
    number = {7424},
    month = {11},
    pages = {431--434},
    volume = {491},
    publisher = {Nature Publishing Group},
    url = {https://www.nature.com/articles/nature11591},
    doi = {10.1038/nature11591},
    issn = {1476-4687}
}

@article{Skillin2024StiffnessAlignment,
    title = {{Stiffness anisotropy coordinates supracellular contractility driving long-range myotube-ECM alignment}},
    year = {2024},
    journal = {Science Advances},
    author = {Skillin, Nathaniel P. and Kirkpatrick, Bruce E. and Herbert, Katie M. and Nelson, Benjamin R. and Hach, Grace K. and G{\"{u}}nay, Kemal Arda and Khan, Ryan M. and DelRio, Frank W. and White, Timothy J. and Anseth, Kristi S.},
    number = {22},
    month = {5},
    pages = {31},
    volume = {10},
    publisher = {American Association for the Advancement of Science},
    url = {https://www.science.org/doi/10.1126/sciadv.adn0235},
    doi = {10.1126/SCIADV.ADN0235/SUPPL{\_}FILE/SCIADV.ADN0235{\_}MOVIES{\_}S1{\_}TO{\_}S3.ZIP},
    issn = {23752548},
    pmid = {38820155}
}

@misc{Parmar2025SupplementalSubstrate,
    title = {{Supplemental Material for “Proliferating nematic that collectively senses an anisotropic substrate”}},
    year = {2025},
    author = {Parmar, Toshi and Brauns, Fridtjof and Luo, Yimin and Marchetti, M. Cristina},
    url = {DOI: https://doi.org/10.1103/5nx2-fqgt}
}

@article{Diaz-de-la-Loza2024TheDriver,
    title = {{The extracellular matrix in tissue morphogenesis: No longer a backseat driver}},
    year = {2024},
    journal = {Cells {\&} Development},
    author = {D{\'{i}}az-de-la-Loza, María del Carmen and Stramer, Brian M.},
    month = {3},
    pages = {203883},
    volume = {177},
    publisher = {Elsevier},
    doi = {10.1016/J.CDEV.2023.203883},
    issn = {2667-2901},
    pmid = {37935283}
}

@article{Saw2017TopologicalExtrusion,
    title = {{Topological defects in epithelia govern cell death and extrusion}},
    year = {2017},
    journal = {Nature 2017 544:7649},
    author = {Saw, Thuan Beng and Doostmohammadi, Amin and Nier, Vincent and Kocgozlu, Leyla and Thampi, Sumesh and Toyama, Yusuke and Marcq, Philippe and Lim, Chwee Teck and Yeomans, Julia M. and Ladoux, Benoit},
    number = {7649},
    month = {4},
    pages = {212--216},
    volume = {544},
    publisher = {Nature Publishing Group},
    url = {https://www.nature.com/articles/nature21718},
    doi = {10.1038/nature21718},
    issn = {1476-4687},
    pmid = {28406198}
}

@article{Maroudas-Sacks2020TopologicalMorphogenesis,
    title = {{Topological defects in the nematic order of actin fibres as organization centres of Hydra morphogenesis}},
    year = {2020},
    journal = {Nature Physics 2020 17:2},
    author = {Maroudas-Sacks, Yonit and Garion, Liora and Shani-Zerbib, Lital and Livshits, Anton and Braun, Erez and Keren, Kinneret},
    number = {2},
    month = {11},
    pages = {251--259},
    volume = {17},
    publisher = {Nature Publishing Group},
    url = {https://www.nature.com/articles/s41567-020-01083-1},
    doi = {10.1038/s41567-020-01083-1},
    issn = {1745-2481}
}

@article{Copenhagen2020TopologicalColonies,
    title = {{Topological defects promote layer formation in Myxococcus xanthus colonies}},
    year = {2020},
    journal = {Nature Physics 2020 17:2},
    author = {Copenhagen, Katherine and Alert, Ricard and Wingreen, Ned S. and Shaevitz, Joshua W.},
    number = {2},
    month = {11},
    pages = {211--215},
    volume = {17},
    publisher = {Nature Publishing Group},
    url = {https://www.nature.com/articles/s41567-020-01056-4},
    doi = {10.1038/s41567-020-01056-4},
    issn = {1745-2481},
    arxivId = {2001.03804}
}

@article{Talbott2022WoundFibrosis,
    title = {{Wound healing, fibroblast heterogeneity, and fibrosis}},
    year = {2022},
    journal = {Cell Stem Cell},
    author = {Talbott, Heather E. and Mascharak, Shamik and Griffin, Michelle and Wan, Derrick C. and Longaker, Michael T.},
    number = {8},
    month = {8},
    pages = {1161--1180},
    volume = {29},
    publisher = {Cell Press},
    doi = {10.1016/J.STEM.2022.07.006},
    issn = {1934-5909},
    pmid = {35931028}
}

\end{document}


\textbf{Video 1}: Simulations of the model with no anisotropy (left), with
anisotropic friction (center) and with aligning field (right) for seeding density $50$ cells$/\rm{mm}^{-2}$. The first row of heatmaps represent cell-substrate order and the black segments are local nematic directors (with their lengths proportional to the magnitude of local order). The second row of heatmaps represent density (units of cells/${\rm{mm}^{-2}}$) and the white segments are local nematic directors (with their lengths proportional to the magnitude of local order). The video shows that at low seeding densities, friction anisotropy and aligning field both produce high alignment to substrate and negligible defect numbers. The value of the friction anisotropy for the middle frames is $\epsilon=0.4$, the value of aligning field for the right-most frames is $\Pi=0.002$ and all other parameters are as noted in the main text.

\textbf{Video 2}: Simulations of the model with no anisotropy (left), with
anisotropic friction (center) and with aligning field (right) for seeding density $150$ cells$/\rm{mm}^{-2}$. The first row of heatmaps represent cell-substrate order and the black segments are local nematic directors (with their lengths proportional to the magnitude of local order). The second row of heatmaps represent density (units of cells/${\rm{mm}^{-2}}$) and the white segments are local nematic directors (with their lengths proportional to the magnitude of local order). The video shows that at intermediate seeding densities, friction anisotropy and aligning field both produce significant alignment to substrate and a few defects. The value of the friction anisotropy for the middle frames is $\epsilon=0.4$, the value of aligning field for the right-most frames is $\Pi=0.002$ and all other parameters are as noted in the main text.

\textbf{Video 3}: Simulations of the model with no anisotropy (left), with
anisotropic friction (center) and with aligning field (right) for seeding density $300$ cells$/\rm{mm}^{-2}$. The first row of heatmaps represent cell-substrate order and the black segments are local nematic directors (with their lengths proportional to the magnitude of local order). The second row of heatmaps represent density (units of cells/${\rm{mm}^{-2}}$) and the white segments are local nematic directors (with their lengths proportional to the magnitude of local order). The video shows that at high seeding densities, friction anisotropy and aligning field both produce little to no alignment to substrate and a many defects. There are also regions of space where there is no local order at all (zero stick lengths). The value of the friction anisotropy for the middle frames is $\epsilon=0.4$, the value of aligning field for the right-most frames is $\Pi=0.002$ and all other parameters are as noted in the main text.